\documentclass[letterpaper,twocolumn,10pt]{article}
\usepackage{usenix2019_v3}

\AtBeginDocument{%
  }

\usepackage{longtable}
\usepackage{booktabs}
\usepackage{xcolor}  
\usepackage{array}
\usepackage{hyperref}
\usepackage{tabularx}
\usepackage{multirow}
\usepackage{colortbl}  
\usepackage{makecell}  
\usepackage{fontawesome5}
\usepackage{graphicx}  
\usepackage{acronym}
\usepackage{listings}
\usepackage{tikz}
\usepackage{tikz-network}
\usepackage{graphics}
\usepackage{soul}
\usepackage{enumitem}
\usepackage{comment}
\usepackage[frozencache,cachedir=.]{minted}
\usepackage{bchart}
\usepackage{xspace}
\usepackage{pgfplots}
\usepackage{tabularray}
\usepackage{tcolorbox}
\usepackage{vcell}
\usepgfplotslibrary{statistics}
\pgfplotsset{compat=1.18}
\usepackage{adjustbox}
\newcolumntype{R}[1]{>{\raggedleft\arraybackslash}p{#1}}
\newcolumntype{L}[1]{>{\raggedright\arraybackslash}p{#1}}

\newboolean{showcomments}
 \setboolean{showcomments}{true}

\ifthenelse{\boolean{showcomments}}
  {\newcommand{\nb}[2]{
  \fbox{\bfseries\sffamily\small\texttt{#1}}
     {\sf\small$\blacktriangleright$\texttt{\textcolor{red}{#2}}$\blacktriangleleft$}
   }
  }
  {\newcommand{\nb}[2]{}
   
  }

\usepackage{xspace}
\usepackage{amsfonts}

\newcommand{\etal}{\textit{et al.,}\xspace}
\newcommand{\ie}{{\textit{i.e.,}}\xspace}
\newcommand{\eg}{\textit{e.g.,}\xspace}
\newcommand{\afl}{\textsc{afl++}\xspace}
\newcommand{\targetarch}{{\sc x86}\xspace}
\newcommand{\clang}{{\sc Clang}\xspace}
\newcommand{\llvm}{{\sc llvm}\xspace}
\newcommand{\wllvm}{{\sc wllvm}\xspace}
\newcommand{\gcc}{{\sc gcc}\xspace}

\newcommand{\inlinecode}[1]{%
  \mintinline[fontsize=\footnotesize{},mathescape, escapeinside=||]{cpp}{#1}%
}

\newcommand{\tbl}[1]{Table~\ref{#1}}
\newcommand{\sect}[1]{\S~\ref{#1}}
\newcommand{\fig}[1]{Figure~\ref{#1}}
\newcommand{\lst}[1]{Listing~\ref{#1}}

\newcommand{\apdx}[1]{Appendix~\ref{#1}}

\acrodef{MCU}{Microcontroller Unit}
\acrodef{ISA}{Instruction Set Architecture}
\acrodef{RTOS}{Real Time Operating System}
\acrodefplural{RTOS}[RTOSes]{Real Time Operating Systems}
\acrodef{OS}{Operating System}
\acrodef{MMIO}{Memory Mapped I/O}
\acrodef{HAL}{Hardware Abstraction Layer}
\acrodef{LPL}{Linux Portable Layer}
\acrodef{WPL}{Windows Portable Layer}
\acrodef{NPL}{Native Portable Layer}
\acrodef{ISR}{Interrupt Service Routine}
\acrodef{CDF}{Cumulative Distribution Function}

\newcommand{\numapp}{18}
\newcommand{\numrtos}{four}

\newcommand{\numbugs}{21}
\newcommand{\systemname}{{\sc Lemix}\xspace}
\newcommand{\elapp}{{\sc LeApp}\xspace}
\newcommand{\freertos}{{\sc FreeRTOS}\xspace}

\newcommand{\toolurl}{\textcolor{red}{\url{https://zenodo.org/records/15611391}}\xspace}

\newcommand{\machiry}[1]{\textcolor{green}{\textbf{Machiry:} #1}}
\newcommand{\davis}[1]{\textcolor{red}{\textbf{Davis:} #1}}
\newcommand{\JD}[1]{\davis{#1}}

\newcommand{\ritvik}[1]{\textcolor{orange}{\textbf{Ritvik:} #1}}
\newcommand{\PA}[1]{\textcolor{brown}{\textbf{PA:} #1}}

\newcommand{\numperapplicationtime}{48\xspace}
\newcommand{\numperfunctiontime}{10\xspace}

\newcommand{\nummorebugsmmiocond}{7\xspace}

\newcommand{\nummorebugsfunctionlevel}{\textcolor{black}{11}\xspace}
\newcommand{\numfunctionleveltotalbugs}{\textcolor{black}{28}\xspace}
\newcommand{\numfunctionleveluniquebugs}{\textcolor{black}{11}\xspace}

\newcommand{\Revision}[2]{%
}

\definecolor{PowderBlue}{RGB}{197,217,240}

\newcommand{\highlighted}[1]{{#1}}
\usepackage{xcolor}
\newcommand{\summary}[1]{
\begin{tcolorbox} [width=1.0\linewidth, colback=blue!07!white, top=1pt, bottom=1pt, left=2pt, right=2pt]
#1
\end{tcolorbox}
}

\begin{document}

\title{\systemname{}: Enabling Testing of Embedded Applications as Linux Applications}

\author{
{\rm Sai Ritvik Tanksalkar}\\
Purdue University
\and
{\rm Siddharth Muralee}\\
Purdue University
\and
{\rm Srihari Danduri}\\
Purdue University
\and
{\rm Paschal Amusuo}\\
Purdue University
\and
{\rm Antonio Bianchi} \\
Purdue University
\and
{\rm James C Davis}\\
Purdue University
\and
{\rm Aravind Kumar Machiry}\\
Purdue University
} 

\maketitle
\begin{abstract}

Dynamic analysis, through rehosting, is an important capability for security assessment in embedded systems software. Existing rehosting techniques aim to provide high-fidelity execution by accurately emulating hardware and peripheral interactions.
However, these techniques face challenges in adoption due to the increasing number of available peripherals and the complexities involved in designing emulation models for diverse hardware.
Additionally, contrary to the prevailing belief that guides existing works, our analysis of reported bugs shows that high-fidelity execution is not required to expose most bugs in embedded software.
Our key hypothesis is that security vulnerabilities are more likely to arise at higher abstraction levels.
%

To substantiate our hypothesis, we introduce~\systemname, a framework enabling dynamic analysis of embedded applications by rehosting them as x86 Linux applications decoupled from hardware dependencies. Enabling embedded applications to run natively on Linux facilitates security analysis using available techniques and takes advantage of the powerful hardware available on the Linux platform for higher testing throughput.
We develop various techniques to address the challenges involved in converting embedded applications to Linux applications.
We evaluated \systemname{} on \numapp{} real-world embedded applications across \numrtos{} RTOSes and found \numbugs{} new bugs, in 12 of the applications and all 4 of the RTOS kernels.
We report that \systemname{} is superior to existing state-of-the-art techniques both in terms of code coverage ($\sim$2X more coverage) and bug detection (18 more bugs).

\end{abstract}


\section{Introduction}
\label{sec:introduction}
%



Society's dependence on low-powered \ac{MCU} based devices (\eg{} IoT devices), has significantly increased, controlling various aspects of our daily lives, including homes~\cite{Alrawi2019SoK:Deployments}, transportation~\cite{al2020intelligence}, traffic management~\cite{soni2017review}, and the distribution of vital resources like food~\cite{prapti2022internet} and power~\cite{o2013industrial}. The adoption of these devices has seen rapid and extensive growth, with an estimated count of over 50 billion devices by the end of 2020~\cite{al-garadi_survey_2020}.
Vulnerabilities in the software controlling these devices have far-reaching consequences~\cite{antonakakis2017understanding, writer_5_2020} due to the pervasive and interconnected nature of these devices, as exemplified by the infamous Mirai botnet~\cite{margolis2017depth} and more recent URGENT/11~\cite{noauthor_urgent11_nodate} vulnerabilities.
It is important to detect such vulnerabilities proactively.
Various works \cite{zaazaa2020dynamic} show that dynamic analysis, especially fuzzing \cite{manes2019art}, is effective at vulnerability detection in web and desktop software.
However, the dynamic analysis of embedded systems \cite{garousi2018testing} is challenging \cite{muench2018you,white_making_2011} because of the close interaction with hardware and the lack of \ac{OS} abstractions.
The lack of robust and readily available dynamic analysis tools (comparable to those for x86 systems) further imposes engineering challenges.

To mitigate this, \emph{rehosting} \cite{fasano2021sok} has emerged as an effective technique.
By decoupling firmware from its hardware dependencies and enabling execution within an emulated environment, rehosting facilitates deeper exploration and analysis of embedded software without the constraints of physical hardware.
Existing rehosting techniques mainly focus on achieving high-fidelity execution without hardware and focus on modeling peripheral interactions through manually created models~\cite{clements_halucinator_2020}, pattern-based model generation~\cite{feng2020p2im}, or models built using machine learning techniques~\cite{spensky2021conware, gustafson2019toward}.
They depend on the availability of an~\ac{MCU}-specific~\ac{ISA} emulator and require considerable engineering effort~\cite{RehostingChallenges} to configure different peripherals.
\emph{We hypothesize that this high-fidelity execution is not required for vulnerability detection, and a coarse approximation of program behavior is sufficient}. 
We validate our hypothesis through a preliminary analysis of previously reported bugs (\sect{subsec:bugmanifestationfidelity}).
We find that most bugs occur at higher software levels, and not within architecture-specific elements like inline assembly.

Starting from this observation, in this paper, we present \systemname{}, a novel approach to rehost embedded applications as Linux applications (for x86), which we call \elapp{}s, with the goal of improving vulnerability detection capability in embedded software with minimal engineering effort.
\systemname{} enables the use of dynamic analysis techniques readily available for Linux applications, such as sanitizers \cite{serebryany2012addresssanitizer} on embedded applications.
However, converting embedded applications to x86 Linux applications and enabling dynamic analysis poses challenges, \ie{} preserving execution semantics, retargeting to different \ac{ISA}, and handling peripheral interactions.
We maintain execution semantics by leveraging the Linux Portable Layer, which comes as a part of most of the prevalent RTOSes (\sect{subsubsec:handlingexecsemantics}).
We use an interactive refactoring approach (\sect{subsubsec:interactivecompilererrorresolution}) to handle \ac{ISA} retargeting.
We tackle peripheral interactions (\sect{subsubsec:periphelmodeling}) by first identifying MMIO addresses through constant address analysis and using runtime instrumentation to feed peripheral data through standard input, thereby eliminating the need for precise peripheral models.
We also weaken peripheral state-dependent conditions to improve code coverage, which is often limited by these conditions that are difficult for a fuzzer to bypass.
To further improve testing, we apply a function-level fuzzing approach based on available research \cite{AFGen, murali2024fuzzslice} that directly invokes the target function with appropriate arguments generated from the input.
\Revision{2. Novelty issue: further clarify the contributions of their approach, especially the novelty part.}{Substantiating novelty claims.}
\highlighted{Taken together, these design choices form a novel rehosting methodology that enables efficient bug discovery in embedded applications without sacrificing practical effectiveness, as demonstrated by our findings.}

We evaluated \systemname{} on \numapp{} real-world embedded applications ranging across \numrtos{} RTOSes, including FreeRTOS, Nuttx, Zephyr, and Threadx. These RTOSes support major semiconductor platforms like Qualcomm, NXP, Nordic~\cite{freertosSemiconductorPartners, apacheSupportedPlatforms, zephyrVendors, threadxvendors} etc.
We show that our approach can successfully convert applications to \elapp{} with only a little manual effort.
We tested \elapp{}s by using whole-program fuzzing and function-level fuzzing and found \numbugs{} previously unknown bugs with 14 out of 18 applications effected by these bugs.
Our ablation study shows that each of our techniques significantly contribute to the overall effectiveness of \systemname{}.
Finally, comparative evaluation against the state-of-the-art approaches shows that \systemname{} is superior at improving code coverage ($\sim$2X more coverage) and bug detection (18 additional bugs).

In summary, the following are our contributions:
\begin{itemize}[noitemsep, leftmargin=*]
\item We propose~\systemname, an extensible framework to rehost embedded applications as x86 Linux applications (\ie{} \elapp{}s) without emulation or physical devices.
\item We design various analysis techniques to tackle challenges in maintaining execution semantics, retargeting, and handling peripheral interactions.
We also design techniques to improve the testing and code-coverage of \elapp{}s.

\item We evaluated \systemname{} on \numapp{} embedded applications across \numrtos{} \acp{RTOS} and found \numbugs{} previously unknown bugs, most of which are confirmed and fixed by the corresponding vendors.
\item Our comparative evaluation against state-of-the-art approaches shows that \systemname{} is superior in code coverage and bug detection.

\end{itemize}



\begin{figure*}[t]
\centering
\includegraphics[width=\linewidth]{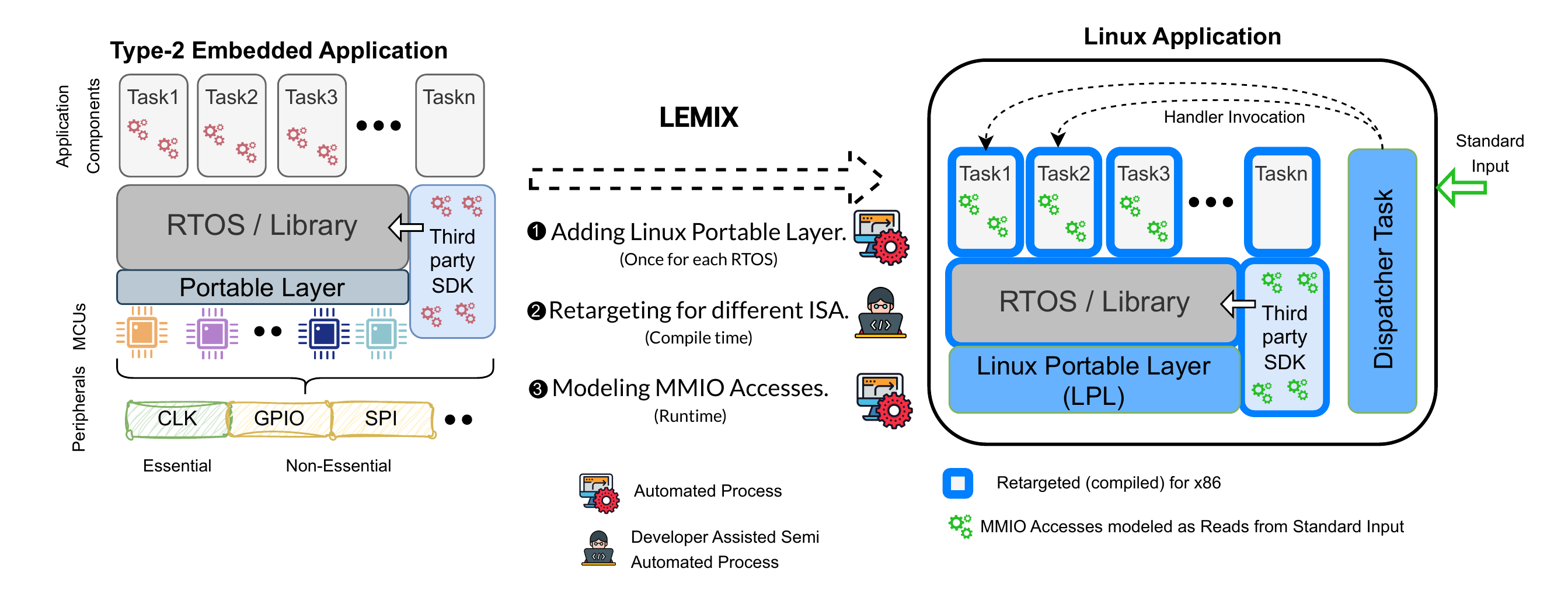}
\caption{
Architecture of a Type-2 Embedded System and overview of \systemname{} approach to convert it to a Linux Application.
}
\vspace{-0.3cm}
\label{fig:background}
\end{figure*}



\section{Background and Threat Model}
\label{sec:background}

\begin{table}[]
\centering
\footnotesize
\begin{tabular}{ccc}
\toprule
\textbf{RTOS}  & \textbf{Low Fidelity} & \textbf{High Fidelity} \\ \toprule
\rowcolor[HTML]{EFEFEF} 
FreeRTOS       & 20                    & 2                      \\ 
Zephyr         & 26                    & 7                      \\
\rowcolor[HTML]{EFEFEF} 
RIOT           & 14                    & 2                      \\ 
\midrule
\textbf{Total} & \textbf{60 (85\%)}           & \textbf{11 (15\%)}            \\ \bottomrule
\end{tabular}
\caption{\highlighted{Our analysis of the CVEs from the Rust4Embedded survey\mbox{~\cite{Rust4EmbeddedSurvey}, \mbox{\cite{sharma2023rust}}} indicates that 60 out of 71 (85\%) require low-fidelity execution. \mbox{~\tbl{tab:embedded_bug_survey}} (Appendix) has category-wise breakdown of the bugs.}}
\label{tab:embedded_bug_survey_mini}
\end{table}

We provide the necessary background of our target embedded systems (\sect{subsec:types_emboss}) and information about their software architecture (\sect{subsec:portability_layers}), along with our threat model (\sect{subsubsec:threatmodel}).

\subsection{Type-2 Embedded Systems}
\label{subsec:types_emboss}
Embedded systems perform a designated task with custom-designed software and hardware. 
Following previous systematization works~\cite{muench2018you,fasano2021sok}, these systems can be categorized into three types:
  Type-1 systems use general purpose \acp{OS} retrofitted for embedded systems,~\eg Embedded Linux;
  Type-2 systems use an \ac{RTOS}, a class of OS that provides timing guarantees, minimal hardware abstraction, and prioritizes tasks to meet strict timing constraints critical for real-time applications,
  and
  Type-3 systems use no \ac{OS} abstractions. 

In this work, we focus on Type-2 systems, which consist of an RTOS combined with application code.
Type-2 designs are common in safety-critical scenarios, supported by the availability of safety-certified \ac{RTOS}es~\cite{vxWorksSafety, safertosCertified, microsoftAzureRtos}, which comply with guidelines like those set by MISRA~\cite{bagnara2018misra} and provide real-time guarantees~\cite{stankovic2004real}.
As shown in~\fig{fig:background}, they have a layered design~\cite{Shen2023NCMAs} and decouple the application components from the underlying \ac{RTOS} kernel.
Most~\ac{RTOS}es modularize their code base to capture all the hardware-specific functionalities within a portability layer specialized per~\ac{MCU}.

\subsection{Portability Layers}
\label{subsec:portability_layers}


As shown in \fig{fig:background}, \acp{RTOS} depend on a portable architecture to enable easy support for the diverse set of available CPU architectures and boards.
Specifically, the portable layer provides header files that define interfaces between the hardware-agnostic kernel and the various \ac{MCU}-specific ports.
The \ac{RTOS} kernel above the portable layer contains hardware-agnostic code.
The hardware-specific implementations, containing interactions with specific \ac{MCU} registers, memory regions, and peripherals, are contained in \ac{MCU} ports, which are compiled and linked with the kernel.
As a result of the portable architecture, an embedded application designed for a specific CPU architecture can run on a different CPU architecture by replacing the current \ac{MCU} port with that of the new architecture~\cite{PortableSoftware}.

To improve testability and aid embedded firmware development, many RTOSes also provide ports for various host operating systems such as Linux and Windows.
We refer to these ports as the \emph{\acf{NPL}} and this includes the \emph{\acf{LPL}} and the \emph{\acf{WPL}}. These native ports allow embedded applications built on these RTOSes to be run on respective desktop operating systems as native applications. 
Native ports use host-provided implementation to simulate various embedded functionalities. {For example, the \protect\acf{LPL} of the FreeRTOS~\cite{FreeRTOS} and Zephyr~\cite{ZephyrRTOS} operating systems use Linux \emph{pthreads} to simulate tasks, \emph{signals} to simulate interrupts, and \emph{timers} to simulate clocks in the application.
We provide more details about \ac{NPL} in \apdx{subsubsec:npldetails}.

\subsection{Threat Model}
\label{subsubsec:threatmodel}

Embedded applications receive inputs from a variety of sources, such as network interfaces, external storage devices (\eg SD cards, USB), user-provided inputs via buttons or screens, and peripherals accessed through \ac{MMIO}.
In our threat model, we assume that \emph{the attacker can control all inputs to the embedded application, including those coming from peripherals accessed via \ac{MMIO} accesses}.
Specifically, all values through \ac{MMIO} reads are fully controlled by the attacker.
The goal of the attacker is to trigger vulnerabilities in the embedded application.

This threat model is reasonable from the Defense in Depth perspective \cite{mughal2018art} and has been used in several other works~\cite{arroyo2021a, ma2022printfuzz, CANNON}.
Also, from a software resilience standpoint, it is important to reasonably validate data received from external entities (such as peripherals) to avoid arbitrary failures.
\Revision{1.3 Present an example that a) shows an actual security impact and b) shows that low-fidelity analysis is sufficient, instead of clarifying that NULL dereferences may not cause issues on all architectures.
}{Added a new infinite loop example in Appendix}
\highlighted{For instance, in \mbox{\lst{lst:mmio_infinite_loop}} (Appendix), blindly trusting the data from MMIO \mbox{\inlinecode{GPIOx->LATCH}} (Line 2) could result in an infinite loop (Line 23), causing DoS.}


\section{Motivation}
\label{sec:motivation}



Dynamic analysis, such as fuzzing, is shown to be an effective technique for vulnerability detection \cite{manes2019art}.
Scalable dynamic analysis of Type-2 embedded applications requires an instrumentation capability (\eg{} through an emulator) and hardware independence.
One of the most popular approaches is~\emph{rehosting}~\cite{fasano2021sok}, where an unmodified embedded firmware will be executed or rehosted in a virtualized environment.
One of the main challenges in rehosting is to achieve execution fidelity.
\Revision{2. Novelty issue: further clarify the contributions of their approach, especially the novelty part.}{The new figure 2 has relevant citations to clearly distinguish Lemix from other works.}
\highlighted{The existing rehosting techniques can be categorized according to the developer/analyst effort and execution fidelity as shown in \mbox{\fig{fig:rehosting_novelty}}.
Ideally, we want to achieve \emph{the highest execution fidelity with the least analyst effort in a hardware-independent manner} --- a known hard problem and the holy grail of rehosting \mbox{\cite{fasano2021sok}}.
Most of the recent rehosting techniques try to achieve high execution fidelity and mainly focus on automated techniques to precisely model peripheral interactions --- which are hard to generalize across peripherals.
Furthermore and more importantly, such high-fidelity execution may not be needed to detect most vulnerabilities.}





\subsection{Execution Fidelity (EF)}
\label{subsec:executionfidelity}
\Revision{3.7: Formal Definition of EF and BMF (Improving Revision)}{More details are added here.}
\highlighted{Adapting\footnote{We build upon the broader categorization of Wright \etal{} by introducing more granular taxonomies, enabling a more detailed assessment of execution fidelity specifically in the context of embedded systems.} the categorization from Wright \mbox{\etal{}}\mbox{\cite{RehostingChallenges}}, Execution Fidelity (EF) in embedded systems can be broadly grouped into four categories: \\
\textit{\textbf{Language Semantic Fidelity (S):}} The degree to which the execution preserves the high-level language (\eg{} C/C++) semantics intended by the programmer, including control flow, data types etc. \\
\textit{\textbf{Assembly Execution Fidelity (A):}} The correctness of executing assembly instructions which constitutes instruction-level behavior and any deviations due to instrumentation etc.
In contrast to \textbf{{S}}, which focuses on high-level program behavior, \textbf{{A}} pertains to low-level execution behavior as specified by the processor’s instruction set architecture (ISA). \\
\textit{\textbf{Peripheral Handling Fidelity (P):}} The extent to which peripheral interactions (\eg{} memory-mapped I/O) are accurately modeled or handled during execution.
While \textbf{A} ensures correct instruction behavior, P focuses on the correctness of effects on peripheral device interaction, requiring hardware modeling beyond the instruction level. \\
\textit{\textbf{Clock Fidelity (C):}} The accuracy of timing behavior with respect to real-time constraints such as instruction timing, interrupts, system clock behavior etc.

For the sake of simplicity, we define Execution Fidelity (EF) as $<S, A, P, C>$, where each component is categorized as \textbf{Low (L)}, \textbf{Medium (M)} or \textbf{High (H)}.
While we adopt this discrete structure for clarity, finer gradations or even a continuous scale (Examples in Appendix \mbox{\sect{apdx:continuousfidelity})} may offer further insights and are left for future work.
This definition of EF also provides a way to categorize existing works.
For instance, hardware-in-the-loop approaches, such as AVATAR \mbox{\cite{zaddach2014avatar}},  redirect all peripheral handling to the real board and execute the embedded firmware on the emulator.
The split execution does not preserve the relative clock semantics between the emulator and actual hardware and only achieves partial clock fidelity, \ie{} C = $M$ (Medium).
The EF achieved by these approaches can be specified as $<H, H, H, M>$.}

\subsection{Bug Manifestation Fidelity (BMF)}
\begin{listing}
    \begin{minted}[xleftmargin=0.25cm, numbersep=1pt, escapeinside=||, fontsize=\scriptsize, breaklines, linenos]{cpp}
    int32_t tud_msc_read10_cb(uint32_t lba, uint32_t offset,
    void* buffer, uint32_t bufsize)
    {    
      // out of ramdisk
      if ( lba >= DISK_BLOCK_NUM ) {
        return -1;
      }
      /* Attacker can offset to sensitive memory */
      uint8_t const* addr = msc_disk[lba] + offset; |\textcolor{orange}{\faUserSecret}|
      /* Controlled write to known memory
         may cause undefined behavior */
      memcpy(buffer, addr, bufsize); |\textcolor{red}{\faBug}|
      return (int32_t) bufsize;
    }
    
\end{minted}
\caption{\highlighted{Lack of bounds check on offset in \mbox{\inlinecode{tud_msc_read10_cb}} allows out-of-bounds read from \mbox{\inlinecode{msc_disk[lba]}}, potentially leading to information disclosure or undefined behavior.}}
\label{lst:motivation_bug_new}
\end{listing}
\highlighted{\emph{BMF} is the minimum fidelity required to reach and observe the effects of the bugs of interest.
BMF varies according to the type of bugs.
For instance, to observe scheduling bugs, we need an accurate clock fidelity, \ie{} C, in addition to the other components, depending on where the bug is.
If scheduling bugs do not involve assembly, we do not need A.
We therefore analyze known vulnerabilities in embedded software to understand the BMF required for memory corruption vulnerabilities (a common class of vulnerabilities).
Specifically, which execution aspects out of \textbf{S, A, P, C} (\mbox{\sect{subsec:executionfidelity}}) are required and which of them can be approximated.}

\subsubsection{Empirical Data}
\label{subsec:embeddedsurvey}
\Revision{1.2 Present results of the evaluation of false positives on the Rust for Embedded Systems study, the one the paper used to justify that low-fidelity is sufficient. To what extent does low-fidelity lead to additional false positives?}{The results are presented in the form of Mini table in main text and elaborate table in appendix}
\highlighted{We manually analyzed 84 publicly reported vulnerabilities in C/C++ software taken from the recent work by Sharma \mbox{\etal{} \cite{rustforembeddedpaper}} to identify what degree of fidelity is required to manifest them.
This included CVEs with available patch information from open-source \mbox{\acp{RTOS}}, \ie{} FreeRTOS, Zephyr, and RIOT.
We considered only the common case of memory corruption vulnerabilities, omitting categories such as weak authentication and SQL injection.
Memory corruption vulnerabilities comprised 71 out of the 84 vulnerabilities.

For each CVE, we identified the target vulnerability and affected function by manually analyzing the CVE description and the corresponding patch.
We then check if the vulnerability can be triggered with low-fidelity rehosting.
Specifically, we target vulnerabilities characterized by an EF of $<H, L, M, M>$, as defined in \mbox{\sect{subsec:executionfidelity}}.
We consider those that meet all requirements to be triggerable with low-fidelity rehosting, else high fidelity is needed.
\mbox{\lst{lst:high_fidelity_cve}} (Appendix) shows an example of a CVE requiring high-fidelity rehosting and \mbox{\lst{lst:low_fidelity_cve}} (Appendix) shows an example of a CVE requiring low-fidelity.
We summarize our results in \mbox{\tbl{tab:embedded_bug_survey_mini}}.
More details can be found in \mbox{\tbl{tab:embedded_bug_survey}} (Appendix).
Our analysis is further confirmed by recent work\mbox{~\cite{tay2023greenhouse}}, which detected various vulnerabilities through low-fidelity dynamic analysis.}

\subsubsection{BMF For Embedded System Software}
\label{subsec:bugmanifestationfidelity}
Based on our empirical study \sect{subsec:embeddedsurvey}, the BMF required for most of the memory corruption bugs is $<H, L, M, M>$, which is what \systemname{} targets.
Following the definitions of Wright \etal{} \cite{RehostingChallenges}, BMF for most memory corruption vulnerabilities can be approximated to module-level execution fidelity.
Specifically, we should be able to execute a module (\ie{} a group of functions) with \emph{enough fidelity} to expose a bug.
\Revision{1.3 Present an example that a) shows an actual security impact and b) shows that low-fidelity analysis is sufficient, instead of clarifying that NULL dereferences may not cause issues on all architectures.
}{A new example added for motivation, added a non-MMIO example here, an MMIO based example can be found for the threat model.}
\highlighted{\mbox{\lst{lst:motivation_bug_new}} demonstrates a motivating example of a bug we discovered in TinyUSB.
The \mbox{\inlinecode{tud_msc_read10_cb}} function lacks bounds checking on the \mbox{\inlinecode{offset}} parameter, allowing out-of-bounds reads from the \mbox{\inlinecode{msc_disk}} array (at line 9), 
which can cause potential information disclosure or even undefined behavior, depending on how the buffer is further used.
We do not need a high-fidelity execution to detect the bug in \mbox{\lst{lst:motivation_bug_new}}.
We just need to execute the function \mbox{\inlinecode{tud_msc_read10_cb}} and pass a large number as \mbox{\inlinecode{offset}}.
We also need the capability to detect out-of-bound memory access (at line 9), which is challenging in embedded systems because of the lack of memory protection mechanisms\mbox{~\cite{muench2018you}}. }
Although we do not require precise peripheral models to trigger the bug, achieving BMF or module-level execution fidelity without them is challenging.
As mentioned in \sect{subsec:types_emboss}, embedded applications are organized into a set of tasks and use a real-time scheduler to trigger the tasks.
To execute the function \inlinecode{tud_msc_read10_cb} in \lst{lst:motivation_bug_new}, we need to ensure the task containing the function gets executed, which further depends on the scheduler, which requires precise models for the clock peripheral.
\emph{Can we achieve BMF without explicitly providing precise peripheral models?}
In summary, we need the capability to execute embedded application, handle MMIO accesses (\ie{} provide data on reads and ignore writes), and detect memory safety violations.

\subsection{The Idea}
\label{subsec:linuxapps}

\begin{figure}
\centering
\begin{tikzpicture}
\def\plotwidth{5}
\def\plotheight{5}
\def\gapwidth{1.3}

\pgfmathsetmacro{\leftx}{(\plotwidth-\gapwidth)/2}
\pgfmathsetmacro{\rightx}{(\plotwidth+\gapwidth)/2}
\pgfmathsetmacro{\bottomy}{(\plotheight-\gapwidth)/2}
\pgfmathsetmacro{\topy}{(\plotheight+\gapwidth)/2}

\fill[left color=red!70!white, right color=green!70!white] 
     (0,\topy - \gapwidth/2) rectangle (\leftx, \plotheight);

\fill[left color=green!70!white, right color=red!70!white] 
     (\rightx,\topy - \gapwidth/2) rectangle (\plotwidth,\plotheight);

\fill[left color=red!70!white, right color=green!70!white] 
     (0,0) rectangle (\leftx,\bottomy + \gapwidth/2);

\fill[left color=green!70!white, right color=red!70!white] 
     (\rightx,0) rectangle (\plotwidth,\bottomy + \gapwidth/2);


\fill[green!70!white] (\leftx,0) rectangle (\rightx,\plotheight); 

\filldraw[black] (1.2,2.1) circle (1pt) node[anchor=west] {\tiny\cite{fuzzware}};
\filldraw[black] (0.5,2.1) circle (1pt) node[anchor=west] {\tiny\cite{feng2020p2im}};
\filldraw[black] (1.2,1.8) circle (1pt) node[anchor=west] {\tiny\cite{FirmAFL}};
\filldraw[black] (0.5,1.8) circle (1pt) node[anchor=west] {\tiny\cite{Em-Fuzz}};

\filldraw[black] (1.2,4.8) circle (1pt) node[anchor=west] {\tiny\cite{li2021library}};
\filldraw[black] (0.5,4.8) circle (1pt) node[anchor=west] {\tiny\cite{Kim2020FirmAETL}};
\filldraw[black] (1.2,4.5) circle (1pt) node[anchor=west] {\tiny\cite{Srivastava2019FirmFuzzAI}};
\filldraw[black] (0.5,4.5) circle (1pt) node[anchor=west] {\tiny\cite{QEMU}};

\filldraw[black] (4.3,2.1) circle (1pt) node[anchor=west] {\tiny\cite{muench_avatar_2018}};
\filldraw[black] (3.6,2.1) circle (1pt) node[anchor=west] {\tiny\cite{li_femu_2010}};
\filldraw[black] (4.3,1.8) circle (1pt) node[anchor=west] {\tiny\cite{ruge_frankenstein_2020}};
\filldraw[black] (4.3,1.5) circle (1pt) node[anchor=west] {\tiny\cite{gui_firmcorn_2020}};
\filldraw[black] (3.6,1.8) circle (1pt) node[anchor=west] {\tiny\cite{corteggiani_inception_nodate}};
\filldraw[black] (3.6,1.5) circle (1pt) node[anchor=west] {\tiny\cite{kammerstetter_embedded_2016}};

\filldraw[black] (3.6,4.8) circle (1pt) node[anchor=west] {\tiny\cite{clements_halucinator_2020}};
\filldraw[black] (3.6,4.5) circle (1pt) node[anchor=west] {\tiny\cite{ChenWBE16}};
\filldraw[black] (4.3,4.8) circle (1pt) node[anchor=west] {\tiny\cite{spensky2021conware}};
\filldraw[black] (4.3,4.5) circle (1pt) node[anchor=west] {\tiny\cite{chen2022metaemu}};

\Text[x=0.95cm,y=1.2cm,fontsize=\footnotesize]{\bf \scriptsize SW Approaches}
\Text[x=0.95cm,y=0.9cm,fontsize=\footnotesize]{\bf \scriptsize With No}
\Text[x=0.95cm,y=0.6cm,fontsize=\footnotesize]{\bf \scriptsize Peripheral}
\Text[x=0.95cm,y=0.3cm,fontsize=\footnotesize]{\bf \scriptsize Modeling}

\Text[x=4.1cm,y=0.8cm,fontsize=\footnotesize]{\bf \scriptsize HW Dependent}
\Text[x=4.1cm,y=0.5cm,fontsize=\footnotesize]{\bf \scriptsize Approaches}

\Text[x=0.95cm,y=3.8cm,fontsize=\footnotesize]{\bf \scriptsize Abstract Model}
\Text[x=0.95cm,y=3.5cm,fontsize=\footnotesize]{\bf \scriptsize of Application}

\Text[x=4.1cm,y=3.8cm,fontsize=\footnotesize]{\bf \scriptsize SW Approaches}
\Text[x=4.1cm,y=3.5cm,fontsize=\footnotesize]{\bf \scriptsize With Precise}
\Text[x=4.1cm,y=3.2cm,fontsize=\footnotesize]{\bf \scriptsize Peripheral}
\Text[x=4.1cm,y=2.9cm,fontsize=\footnotesize]{\bf \scriptsize Modeling}

\node[font=\bf\footnotesize] at (2.5,2.5) {\systemname};
\node[font=\bf\scriptsize] at (2.5,2.2) {(BMF)};

\draw[thick,-] (0,0) -- (5,0) node[anchor=north west] {\bf \tiny high};
\node[font=\bf\tiny] at (0,-0.2) {least};
\node[font=\footnotesize] at (2.6,-0.2) {Execution Fidelity};
\node[font=\footnotesize, rotate=90] at (-0.2,2.6) {Analyst / Developer Effort};
\draw[thick,-] (0,0) -- (0,5) node[anchor=south east] {\bf \tiny high};
\end{tikzpicture}
\caption{\highlighted{Existing Rehosting Techniques v/s \systemname{}}}
\label{fig:rehosting_novelty}
\end{figure}
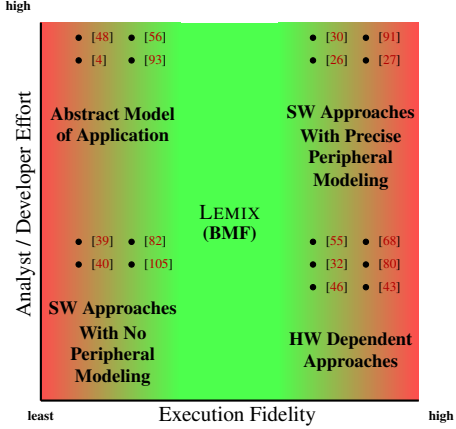

Dynamic analysis challenges like execution environment and detectability have been well studied for Linux applications on standard \acp{ISA} (e.g., x86, x64), with many effective solutions~\cite{d2019sok,song2019sanitizing,bohme2021fuzzing}.
Prior work, such as AoT~\cite{10.1145/3551349.3556915}, extracts components from complex systems (e.g., Linux kernel) into testable user-space applications.
Our goal is to convert embedded applications into Linux applications to enable BMF and make existing dynamic analysis techniques~\cite{FuzzingSurvey} applicable.
Srinivasan \etal{}~\cite{srinivasan2023towards} recently showed this is feasible by manually converting three simple \freertos{} applications.
However, designing a generic technique involves tackling the following challenges.
\begin{itemize}[leftmargin=*]

\item\textbf{(Ch1) Preserving Execution Semantics.} Linux applications, by default, follow single-threaded execution. However, embedded applications (as explained in~\sect{subsec:types_emboss}) are engineered in terms of event-driven tasks and are multithreaded~\cite{henzinger2006embedded}.
Simply replacing RTOS files with their POSIX equivalents (LPL) often leads to unintended errors during integration. Incorporating a POSIX-compatible RTOS requires a systematic and automated mechanism.
This involves more than just file replacements; it necessitates careful adaptation to preserve the embedded system's original task-based and event-driven execution semantics.

\item\textbf{(Ch2) Retargeting to different~\acp{ISA}.} 
Though majorly developed in C, embedded applications use various non-standard and embedded toolchain-specific C features not supported by traditional compilers for desktop \acp{ISA}, \eg{} x86.
The presence of inline \ac{ISA}-specific assembly (\eg{} of ARM) further complicates retargeting (\ie{} compiling) for other \acp{ISA}.
We need to have a mechanism to compile an embedded application for common desktop~\acp{ISA}.
\item\textbf{(Ch3) Handling Peripheral Interactions.}
Embedded systems directly interact with peripherals, mostly through a dedicated set of \ac{MMIO} addresses~\cite{reilly2003memory}.
It is crucial to distinguish these MMIO addresses from regular memory accesses because they correspond to physical hardware components, and improper handling can lead to incorrect behavior.
\end{itemize}
\section{\systemname{}}

\begin{figure*}[t]
\centering
\tiny
\includegraphics[scale=0.4]{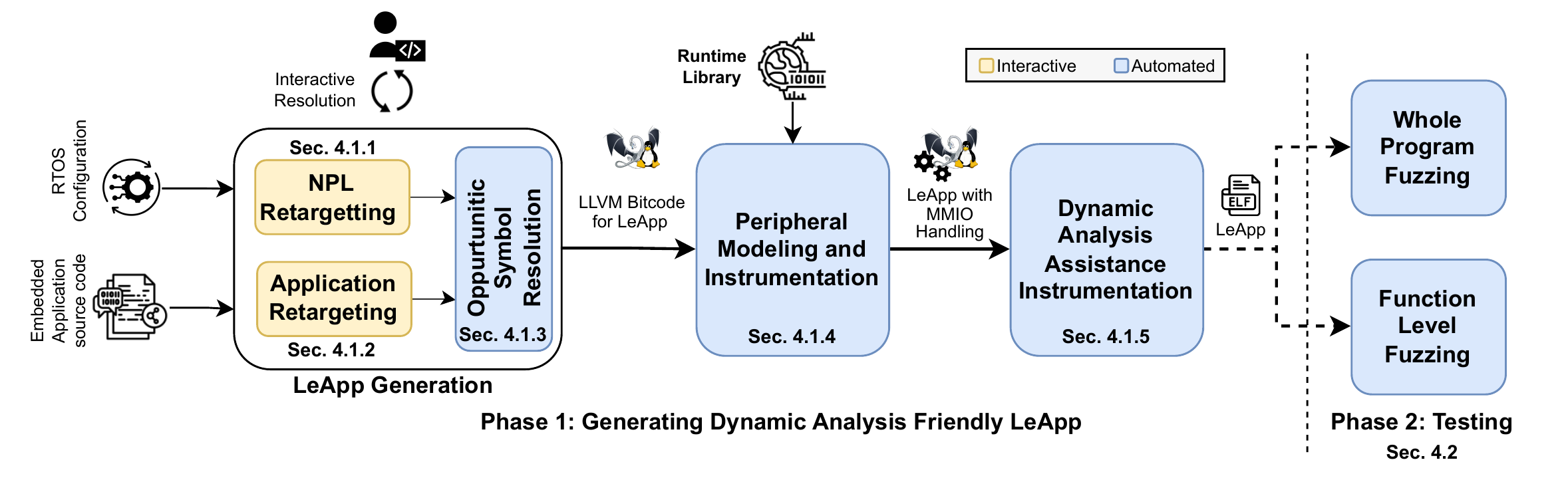}
\caption{
  \highlighted{Overview of our \systemname{} framework.}}
\label{fig:overview_new}
\end{figure*}


We design \systemname{}, an interactive framework enabling effective dynamic analysis of embedded applications by converting them to Linux applications, which we call \elapp{}.
\Revision{2. Novelty issue: further clarify the contributions of their approach, especially the novelty part.}{Substantiating novelty claims.}
\highlighted{The novelty of \systemname{} lies in recognizing and harnessing the BMF insight, \ie{} rehosting with just enough execution fidelity to keep most security bugs triggerable, thereby significantly reducing the manual effort and complexity typically associated with generalizing full system emulation.}
The right side of \fig{fig:background} shows the summary of our approach to tackling the challenges (\sect{subsec:linuxapps}) in converting to \elapp{}.
\Revision{3.2 Move Figure 6 and relevant information from Appendix to main text.}{It is now Figure 3}
\highlighted{The \systemname{} framework has two phases as illustrated in \mbox{\fig{fig:overview_new}}.}
In Phase 1, we convert the given embedded application into \elapp{} using static analysis techniques and compiler instrumentation.
We use an interactive approach to tackle certain complex code idioms during retargeting.
We also design instrumentation techniques for \elapp{}s to improve the effectiveness of dynamic analysis, specifically random testing.
In Phase 2, we focus on testing \elapp{}. We support two modalities, whole-program, and function-level testing, providing a holistic testing infrastructure.

\subsection{Phase 1: Analysis Friendly \elapp{}}
This phase generates a dynamic analysis-friendly \elapp{} from a given embedded application and the target \ac{RTOS} configuration.
This part tackles challenges (1-3) from \sect{subsec:linuxapps}.

\subsubsection{Handling execution semantics using \ac{LPL} (Ch 1)}
\label{subsubsec:handlingexecsemantics}
As explained in \sect{subsec:types_emboss}, embedded applications rely on \ac{RTOS} functions for their execution semantics.
For instance, an application for \freertos{} uses \inlinecode{xTaskCreate} function to create a task and \inlinecode{vTaskStartScheduler} to start the scheduler.
Similarly, \inlinecode{xTimerCreate} function is used to register for a timer event.

Embedded systems use a portable layer enabling an \ac{RTOS} to be used for different \acp{MCU} (\fig{fig:background}).
As explained in \sect{subsec:portability_layers}, most \acp{RTOS} maintain an \ac{NPL} enabling them to run on top of regular OSes, \ie{} Windows (WPL) or Linux (\ac{LPL}).

Given the source code of an embedded application, we identify the \ac{RTOS} dependencies and re-configure them with corresponding \ac{LPL}.
To aid our process, we gather and maintain the \ac{LPL} software packages of \acp{RTOS} apriori.
This is not trivial as the build setup of the application may include \ac{MCU}-specific configurations enabling certain HAL-specific APIs necessary for its functionality.
For instance, in the FreeRTOS app TinyUSB, the \inlinecode{configTIMER_QUEUE_LENGTH} is set to 32, while the POSIX build sets it to 20, causing undefined behavior due to the application's expectation that this value should not exceed its configuration.
In few cases, peripheral models implemented in the original \ac{RTOS} may not be available in the corresponding \ac{LPL}.

To address this, we designed a fully automated approach that selectively integrates configurations from the application that do not disrupt the \ac{LPL} build.
Each \ac{RTOS} configuration from the application is iteratively toggled in the \ac{LPL} build, retaining those that compile successfully.
Once the \ac{LPL} build successfully incorporates the necessary configurations, we replace the application's \ac{RTOS} object files with those from the successful \ac{LPL} build.
We term this process as \textbf{POSIX Swap}.
This approach however can induce unexpected behaviors in the ported application since not every configuration was incorporated from the application's config.
But, we did not observe any false positive crashes due to misconfigured \ac{LPL} build during our evaluation.
 
\subsubsection{Interactive Resolution for Retargeting (Ch 2)}
\label{subsubsec:interactivecompilererrorresolution}
As mentioned in \sect{subsec:linuxapps}, our goal is to build \elapp{} for common desktop \acp{ISA}, specifically x86, because of the availability of various testing tools.
We want to use the \clang{} compiler as the LLVM IR enables us to easily perform various analysis tasks, and also, several techniques (\eg{} loop analysis) already exist in the \clang{} infrastructure.
However, just replacing the compiler with \clang{} and changing the target \ac{ISA} to x86 does not work.
Because (as mentioned in \sect{subsec:linuxapps}) embedded applications use non-standard C language features and inline assembly of other \acp{ISA}, \eg{} ARM.
Handling this requires program semantic reasoning \cite{lim2019automatic}, a known hard problem.

We use an \emph{interactive human-in-the-loop refactoring approach to tackle this}.
We aim to automatically refactor the code to be \clang{} and x86 friendly using a set of refactor patterns.
However, for cases requiring semantic reasoning, we resort to developer assistance by providing precise guidance instructions.
Our automation takes over after developer assistance, and the process continues with intermittent manual refactorings until the resulting code can be compiled using \clang{}, \ie{} able to generate LLVM Bitcode.
The \tbl{tab:gcc_clang_diffs} (Appendix) summarizes automated and interactive refactorings.
\Revision{3.4 Add reference to build process tracing methodology (Appendix B.2) in main text.
}{Done here}
\highlighted{Further details of the build process tracing and streamlining the build system for x86-clang can be found in \mbox{\apdx{apdx:subsecimpl}}.}

We classify the set of refactorings into the following two categories and present the techniques used to handle them:

\begin{enumerate} [leftmargin=*, noitemsep]
\item\textbf{Compiler Incompatibilities:}
These are incompatibilities because of compilers (\gcc{} v/s \clang{}) and architecture-dependent code, \eg{} expecting \inlinecode{int} to be of 4 bytes.
A significant portion of embedded software relies on \gcc-based toolchains~\cite{von2011definitive}.
Hence, making the transition from a \gcc{} build environment to \clang{} is challenging, especially for embedded codebases \cite{Shen2023NCMAs}.
\tbl{tab:gcc_clang_diffs} (Appendix) highlights the incompatibilities between \gcc{} and \clang{} affecting our embedded applications.
Some of these, such as Variable-Sized Object initialization, are still not supported even in the latest version of LLVM at the time of writing (LLVM 18)~\cite{dgookin_2023}.
Although several works \cite{difuze, Shen2023NCMAs, drchecker} mention this problem, to the best of our knowledge, we are the first to highlight these issues, which have not yet received sufficient attention among embedded developers.
Addressing compiler incompatibilities requires semantically equivalent refactorings.
We define a set of refactoring templates for automatically handling several of these issues and resort to developer assistance for others.
We also provide guidance instructions to assist in the refactoring an example of which is shown in Listing~\ref{lst:lemix_guidance} (Appendix).

\vspace{0.25cm}
\item\textbf{Inline Assembly:}
Embedded applications often use inline assembly for low-level operations, such as \ac{MCU}-specific initialization~\cite{Shen2023NCMAs}. \elapp{} eliminates the need for such initialization by relying on \ac{LPL}.
As discussed in \sect{subsec:bugmanifestationfidelity}, precise handling of assembly is unnecessary for manifesting most vulnerabilities. 
We automate source code rewriting to identify and comment out inline assembly regions. Commenting out assembly may lead to uninitialized or undefined variables (\eg{} \lst{lst:inlineASM} Appendix).
Most inline assembly reads architecture-specific registers for initialization checks.
To address this, we randomly initialize variables defined by assembly to 0 or 1, allowing applications to bypass initialization routines over multiple runs.
Our approach may not handle all cases, such as inline assembly within macros or machine code representations (\eg{} \lst{lst:hexASM} in Appendix).
In such cases, we automatically detect issues and provide developers with precise instructions, such as resolving compilation errors like \textit{expected closing parenthesis}.
\systemname{} pinpoints problematic lines and suggests fixes, ensuring minimal manual intervention.

\end{enumerate}




\subsubsection{Opportunistic Symbol Resolution}
\label{subsubsec:oppsymbolresolution}
\Revision{3.6 Clarify that proposed approach incrementally resolves linker errors by selectively linking only the required object files from the prebuilt LPL, resolving linker conflicts in an automated fashion.}{This clarification has been made here.}
The final step in creating the \elapp{} involves linking compiled \ac{LPL} and application object files. However, directly linking RTOS object files often results in linker errors~\cite{10.1145/3375894.3375897} because embedded applications may invoke MCU-specific functions~\cite{Shen2023NCMAs} that are not present in \ac{LPL}, causing undefined reference errors~\cite{undefinedreference}. For example, in the FreeRTOS application \textit{Infinitime}, the function \textit{xTimerGenericCommand} is invoked but not available in \ac{LPL}, leading to a linker error.

\highlighted{\systemname{} incrementally resolves linker errors by selectively linking only the required object files from the prebuilt LPL, resolving linker conflicts in an automated fashion.} This is achieved using an opportunistic approach by identifying and linking MCU-specific RTOS object files (denoted as $O_{u}$) that define the missing symbols.
However, this can cause multiple definition errors if symbols in $O_{u}$ overlap with those in \ac{LPL}.
For instance, resolving \textit{xTimerGenericCommand} by including the application kernel's \textit{timers.o} introduces duplicate definitions, such as \textit{prvInitialiseNewTimer}, between the application kernel's \textit{timers.o} and the \ac{LPL} kernel's \textit{timers.o}.
We want to ensure that references are linked with the expected symbols.
We use a two-phase approach to tackle this:

\begin{enumerate}[leftmargin=*, noitemsep]
    \item\textbf{Creating Library Archives:} 
    We observed that embedded applications and \acp{RTOS} use build procedures based on archiving \cite{archivear}, which prevents duplicate symbol errors across multiple components, \eg{} an application can have a function (say $f$) with the same function as an \ac{RTOS} function. 
Creating an archive of application-specific object files ensures that all references to $f$ within these object files are linked to the application-specific version.
We trace the build process to identify which archives (and their order) were created and the corresponding object files.
We follow the same order when creating and linking archive files to ensure that original references are intact.
\vspace{0.25cm}
    \item\textbf{Modifying Symbol in One of the Objects:} The remaining multiple references cannot be resolved by archiving; hence, the symbol name has to be modified in one of the object files.
If the multiple reference is between an Application's object file and \ac{LPL} object file, we change the symbol name in the Application object file, which ensures that calls in the application to \ac{LPL} functions are linked appropriately.
\end{enumerate}
\noindent\textbf{Handling Symbol Aliasing:}
In the original application's build procedure, the linker might create aliases for various symbols to maintain interoperability across different boards as guided by linker scripts.
For instance, the symbol \textit{\_\_init\_clock} might be resolved to aliases like \textit{\_\_stm32\_clock\_init} or \textit{\_\_nrf52\_clock\_init}, depending on the target board.
Additionally, multiple aliases can be created for the same symbol.
Ignoring such aliases results in \inlinecode{NULL}-ptr deferences while executing the resulting \elapp{}.
\Revision{3.5 Explain the symbol modification process, including alias analysis via firmware debug symbols.}{Done here.}
\highlighted{To tackle this, we first extract the aliasing information using firmware debug symbols from the original embedded application (compiled for the target \mbox{\ac{MCU}}) and identify symbol aliases with the help of GDB.
Next, we instrument our \elapp{} and link the symbol aliases to the appropriate references in \elapp{}.}

\subsubsection{Peripheral Modeling and Instrumentation (Tackling Ch 3)}
\label{subsubsec:periphelmodeling}

As mentioned in \sect{subsubsec:threatmodel}, our threat model includes malicious peripherals, \ie we assume that all inputs from peripherals can be controlled by an attacker.
As mentioned in \sect{subsec:types_emboss}, applications interact with peripherals by accessing corresponding \ac{MMIO} addresses and interrupts.

\noindent\textbf{Handling MMIO Accesses:}
Input from peripherals is received through reading  \ac{MMIO} addresses.
We aim to model loads (\ie{} reads) from these addresses as reads from standard input and ignore stores (\ie{} writes) as we focus on vulnerability detection (\ie{} BMF as described in \sect{subsec:bugmanifestationfidelity}).

First, we determine \ac{MMIO} address ranges. 
One of the common techniques is to find these address ranges from peripheral System View Description (SVD) files \cite{SVDFormat}.
However, as we will show in \sect{subsubsec:peripheraleval}, oftentimes, SVD files are incomplete and do not contain all peripheral address ranges which is also a known problem~\cite{SVDMissing}.
We aim to create an automated technique that does not depend on SVD files.
Peripherals have predefined \ac{MMIO} address ranges, and applications access them through hardcoded addresses~\cite{spensky2021conware, feng2020p2im}.
We perform a constant address analysis to determine all hardcoded address values, \ie{} constant values used as pointer operands in load and store instructions.
The corresponding pages form the base \ac{MMIO} pages ($P_{m}$).
For instance, if we found a constant address $x$, then we will add the corresponding page $[b, b+4093]$ to $P_{m}$, where $b = (x\&\sim{}(0x3FF))$ is the base address of the corresponding page.
We also perform additional coalescing and consider all pages within a range of ±2 KB from that boundary, also as \ac{MMIO} addresses. This approach helps group related \ac{MMIO} access and ensures that accesses within the same memory-mapped region are consistently recognized.

Next, we will hook all loads and stores through compile-time instrumentation and link with our runtime library.
At runtime, our hook will check if a load is within an \ac{MMIO} address range; if yes, then it will read an appropriate number of bytes from input and return the corresponding value.
Similarly, our hook will ignore all stores to \ac{MMIO} address ranges.
\lst{lst:mmio_instrumentation} shows the example of our instrumentation (highlighted lines), where the memory access \inlinecode{RCC->PLLCFGR} is checked to see if it is an MMIO address; if yes, we will read a value of corresponding size (\ie{} 4 bytes) from input.

\noindent\textbf{Handling Interrupts:}
Interrupts are treated as peripheral inputs and triggered at random intervals.
\systemname{} identifies \acp{ISR} using \ac{RTOS}-specific patterns, such as ISR vector tables in assembly files, while ignoring handlers implemented in assembly. Using \ac{RTOS}-specific templates, we create a \emph{Dispatcher Task} to invoke \acp{ISR} at arbitrary intervals (\lst{lst:isr_template} in the Appendix). To prevent false crashes from \acp{ISR} requiring preconditions (e.g., valid global pointers), we use lightweight binary analysis and dynamic tracing to identify and disable them.

\begin{listing}
    \begin{minted}[xleftmargin=0.25cm, numbersep=1pt, escapeinside=||, fontsize=\scriptsize, breaklines, highlightlines={4-6,8}, linenos]{cpp}
    uint32_t HAL_RCC_GetSysClockFreq(void) {
      uint32_t pllm = 0U, pllvco = 0U, pllp = 0U;
      uint32_t sysclockfreq = 0U;
      if (isMMIO(RCC->PLLCFGR)) {
      pllm = get_input_from_stdin() & RCC_PLLCFGR_PLLM;
      } else {
      pllm = RCC->PLLCFGR & RCC_PLLCFGR_PLLM; |\textcolor{blue}{\faMinusSquare}|
      }
      ...    
      sysclockfreq = pllvco / pllp; |\textcolor{red}{\faBug}| ...}
\end{minted}
\caption{\textcolor{blue}{\faIcon{minus-square}} shows original MMIO accesses that are instrumented in \elapp{}.~\textcolor{red}{\faIcon{bug}} shows a div-by-zero bug in STM32.}
\label{lst:mmio_instrumentation}
\end{listing}

\vspace{-0.3cm}
\subsubsection{Dynamic Analysis Assistance Instrumentation}
\label{subsubsec:dynamicanalysisassistance}
Embedded applications have considerable peripheral state-dependent code \cite{fuzzware}.
Specifically, they check for peripherals to be in a specific state before interacting with it or to perform some interesting function, \eg{} as shown in \lst{lst:blocker}, at lines 3-4, the code busy-waits until the control register (accessed through MMIO read \inlinecode{NRF_CLOCK->LFCLKSTAT}) has any of the bits corresponding to \inlinecode{CLOCK_LFCLKSTAT_STATE_Pos} that are not set.

Peripherals state is accessed through reading certain registers  \cite{lei2024friend, fuzzware}, \eg{} clock state is accessed through \inlinecode{NRF_CLOCK->LFCLKSTAT} (an MMIO address) in \lst{lst:blocker}.
Since we model all peripheral reads (\sect{subsubsec:periphelmodeling}) as reads from standard input, the coverage of state-dependent code becomes the problem of constraint input generation.
For instance, in \lst{lst:blocker}, the MMIO access, \ie{}, \inlinecode{NRF_CLOCK->LFCLKSTAT} will be fetched from input.
For the execution to reach out of the loop, an input generation technique (\eg{} \afl{}) should provide an input that satisfies the constraint.
Existing techniques handle this by providing precise peripheral models \cite{fuzzware, spensky2021conware} or symbolic execution \cite{davidson2013fie}, but they have scalability issues \cite{fasano2021sok}.

To tackle this, we perform \emph{Weakening of Peripheral State Dependent conditions}.
Specifically, we instrument each conditional instruction to check whether it involves reading from an MMIO address; if yes, we weaken the condition such that any value can satisfy the constraint with 50\% probability as shown in the lower part of \lst{lst:blocker}.
Previous works \cite{peng2018t, choi2019grey} show that such an approach improves the effectiveness of fuzzing.
We will also show in \sect{subsec:ablationstudy} that our approach greatly improves the coverage.
We also perform instrumentation to collect additional metrics, such as coverage.

\begin{listing}
    \begin{minted}[xleftmargin=0.25cm, numbersep=1pt, escapeinside=||, fontsize=\scriptsize, breaklines, linenos]{cpp}
    // Before Condition Weakening
    while 
        ((NRF_CLOCK->LFCLKSTAT & CLOCK_LFCLKSTAT_STATE_Pos)|\textcolor{red}{\faBan}|) {
        // Busy Waiting
        |\textcolor{orange}{\faSpinner}| }; 
    interesting_function();
    // After Condition Weakening
    label:
        bool cond = (NRF_CLOCK->LFCLKSTAT & 
                        CLOCK_LFCLKSTAT_STATE_Pos); |\textcolor{red}{\faBan}|
        new_cond = cond;
        if (isMMIO(NRF_CLOCK->LFCLKSTAT) && stdin_read() % 2) |\textcolor{blue}{\faWrench}|
            new_cond = !cond; |\textcolor{black}{\faToggleOff}|
        while (new_cond) {goto label};
    interesting_function();
\end{minted}
\caption{\textcolor{red}{\faBan} indicates MMIO coverage blockers, \textcolor{orange}{\faSpinner} marks busy waiting due to unsolved constraints, \textcolor{blue}{\faWrench} represents our instrumentation and \textcolor{black}{\faToggleOff} shows MMIO condition toggling.}
\label{lst:blocker}
\end{listing}

\vspace{-3pt}
\subsection{Phase 2: Testing}
\label{subsec:lemixtesting}
This phase focuses on fuzzing of \elapp{}s generated in Phase 1.
We explore two modes of fuzzing: (i) Whole program and (ii) Function level.
\subsubsection{Whole Program Fuzzing}
\label{subsubsec:wholprogramfuzz}
Here, we fuzz \elapp{}s as a whole by providing inputs at appropriate locations (\ie{} MMIO accesses) until \elapp{} terminates or crashes because of a bug.

\vspace{-4pt}
\subsubsection{Function Level Fuzzing}
\label{subsubsec:functionlevelfuzz}


In this mode, we directly fuzz individual functions by providing arguments of appropriate type \cite{zhangintelligen, 10.1145/3338906.3340456}.
Given a function $f$, we use a simple co-relation analysis \cite{pratikakis2006locksmith} to determine the argument types and their size associations, \eg{} for a function that accepts two arguments, an integer array pointer, and its length; our co-relation analysis will produce: \inlinecode{{arg1: {ARRAY, int, SIZE: arg2}, arg2: int}}.
Next, we automatically create generators for each of the argument types, \ie{} functions that generate values of a specific type from the input.
For instance,  for the above example, the generator will create an integer array of an arbitrary size, populate it with random integer values, and return the pointer and size.
Finally, we invoke $f$ with the pointer returned by the generator and the size as the second argument.
Unlike whole program fuzzing (\sect{subsubsec:wholprogramfuzz}), each fuzzing run in function level fuzzing invokes the target function once and exits.


\section{Evaluation}
\label{sec:evaluation}

\setlength{\tabcolsep}{4pt}
\begin{table}
\tiny
 \caption{Approximate Number of unique basic blocks discovered by various configurations of~\systemname{} in comparison to State Of The Art Tools (Discussed in \sect{subsec:testingelapp} and \sect{subsec:comparativeeval}).
 M1- M3 represents different configuration modes for~\systemname{}. Refer to~\tbl{tab:tool_eval_large} (Appendix) for a larger version.}
\vspace{0.25cm}
\centering
\rowcolors{5}{gray!15}{white}
\begin{tabular}{c | c c | c c | c c | c c c | c c c}
\toprule
& \multicolumn{6}{c|}{\textbf{Lx}} & \multicolumn{3}{c|}{\multirow{2}{*}{\textbf{Fw}}} & \multicolumn{3}{c}{\multirow{2}{*}{\textbf{Mf}}} \\
\cmidrule{2-7}
\multirow{-3}{*}{\textbf{AppID}} & 
\multicolumn{2}{c|}{\textbf{M1}} & \multicolumn{2}{c|}{\textbf{M2}} & \multicolumn{2}{c|}{\textbf{M3}} & 
\multicolumn{3}{c|}{\textbf{}} & 
\multicolumn{3}{c}{\textbf{}} \\ 
\cmidrule{2-7} \cmidrule{8-13}
& Cov & Bug & Cov & Bug & Cov & Bug & Cov & Crash & Bug & Cov & Crash & Bug \\
\midrule
f1 & 731 & 0 & 2.9k & 1 & \textcolor{red}{\faCode} & 1 & 500 & 1 & 0 & 1k & 0 & 0 \\
\hline
f2 & 456 & 1 & 2.8k & 3 & \textcolor{red}{\faCode} & 1 & \textcolor{orange}{\faExclamationCircle} & N/A & N/A & 1.5k & 1 & 1 \\ 
\hline
f3 & 560 & 1 & 668 & 2 & 1.5k & 3 & \textcolor{orange}{\faExclamationCircle} & N/A & N/A & \textcolor{orange}{\faExclamationCircle} & N/A & N/A \\
\hline
f4 & 563 & 0 & 1k & 1 & 6k & 3 & 500 & 0 & 0 & 2k & 41 & 0 \\
\hline
f5 & 442 & 0 & 728 & 2 & 1.8k & 2 & 700 & 0 & 0 & 1.8k & 93 & 0 \\
\midrule
n1 & 105 & 0 & 301 & 1 & 13.5k & 1 & 356 & 0 & 0 & 25.2k & 148 & 0 \\
\hline
n2 & 143 & 0 & 338 & 0 & 16.8k & 1 & 405 & 0 & 0 & 300 & 0 & 0 \\
\hline
n3 & 157 & 1 & 357 & 1 & 19.9k & 1 & 60 & 1 & 1 & 100 & 1 & 0 \\
\hline
n4 & 135 & 0 & 235 & 1 & 19.5k & 1 & 388 & 3 & 0 & 2.4k & 0 & 0 \\
\hline
n5 & 823  & 0 & 1.5k & 1 & \textcolor{red}{\faCode} & 0 & 134 & 0 & 0 & 226 & 0 & 0 \\
\midrule
z1 & 76 & 0 & 86 & 2 & 3k & 4 & 44 & 2 & 0 & 213 & 0 & 0 \\
\hline
z2 & \textcolor{blue}{\faMicrochip} & 0 & \textcolor{blue}{\faMicrochip} & 2 & 12.3k & 3 & \textcolor{orange}{\faExclamationCircle} & N/A & N/A & 10.8k &  483 & 1 \\
\hline
z3 & \textcolor{blue}{\faMicrochip} & 0 & \textcolor{blue}{\faMicrochip} & 2 & 4.6k & 6 & \textcolor{orange}{\faExclamationCircle} & N/A & N/A & \textcolor{orange}{\faExclamationCircle} & N/A & N/A \\
\midrule
t1 & 210 & 0 & 553 & 0 & 6.3k & 0 & 670 & 0 & 0 & 750 & 0 & 0 \\
\hline
t2 & 214 & 0 & 553 & 0 & 240 & 0 & 791 & 0 & 0 & 694 & 0 & 0 \\
\hline
t3 & \textcolor{blue}{\faMicrochip} & 0 & \textcolor{blue}{\faMicrochip} & 0 & 3.1k & 0 & 900 & 0 & 0 & 700 & 0 & 0 \\
\hline
t4 & 310 & 0 & 455 & 1 & 188 & 1 & 749 & 0 & 0 & 659 & 10 & 1 \\
\hline
t5 & 254 & 0 & 436 & 0 & 5.2k & 0 & 748 & 0 & 0 & 658 & 0 & 0 \\ \midrule
Avg/Tot & 345 & 3 & 860 & 20 & 7.5k & 28 & 496  & 7 & 1 & 3.1k & 777 & 3 \\ 
\midrule
Unique Bugs &  & 3 &  & 10 &  & 11 &  &  & 1 &  &  & 3 \\ 

\bottomrule
\end{tabular}
\label{tab:tool_eval}
\end{table}

We use a combination of Python scripts and \clang{}/\llvm{} 10 toolchain passes to implement our framework.
We provide more details in \apdx{apdx:subsecimpl}.
We evaluate \systemname{} by answering the following questions:
\begin{enumerate}[noitemsep]
\item[\textbf{RQ1}] (\textit{Converting to \elapp{} (\sect{subsec:rq1eval})}): How effective is our approach (\sect{subsubsec:interactivecompilererrorresolution}) in converting embedded applications to \elapp{}s? How much manual effort does it require?
\item[\textbf{RQ2}] (\textit{Peripheral Handling (\sect{subsubsec:peripheraleval})}): How effective is our approach (\sect{subsubsec:periphelmodeling}) in identifying~\ac{MMIO} addresses?
\item[\textbf{RQ3}] (\textit{Testing~\elapp{} Applications (\sect{subsec:testingelapp})}): What is the effectiveness of testing~\elapp{}s through different fuzzing approaches,~\ie{} whole-program fuzzing and function-level fuzzing? 
\item[\textbf{RQ4}] (\textit{Ablation Study (\sect{subsec:ablationstudy})}): What is the contribution of our peripheral handling (\sect{subsubsec:periphelmodeling}) and dynamic analysis assistance (\sect{subsubsec:dynamicanalysisassistance}) on overall effectiveness?
\item[\textbf{RQ5}] (\textit{Comparative Evaluation (\sect{subsec:comparativeeval})}): What is the effectiveness of \systemname{} compared to the existing state-of-the-art?
\Revision{1.1. Test whether low-fidelity analysis lead to more false positives than competing approaches.}{The FP Section has details about our FP analysis}
\item[\textbf{RQ6}] \highlighted{(\textit{False Positive Analysis (\mbox{\sect{subsec:falsepositives}}})): What false positives are introduced by low-fidelity execution, how do they compare against existing State-Of-The-Art, and how are they remediated?}
\end{enumerate}

\subsection{Dataset and Setup}
\label{subsec:datasetset}

\begin{table}
\scriptsize
\centering
\begin{tblr}{
  row{even} = {c},
  row{3} = {c},
  row{5} = {c},
  row{7} = {c},
  row{9} = {c},
  row{11} = {c},
  row{13} = {c},
  row{15} = {c},
  row{17} = {c},
  row{19} = {c},
  cell{1}{1} = {c},
  cell{1}{2} = {c,t},
  cell{1}{3} = {c,t},
  cell{1}{4} = {t},
  cell{1}{5} = {t},
  cell{1}{6} = {t},
  cell{2}{1} = {r=5}{},
  cell{5}{6} = {r=2}{},
  cell{7}{1} = {r=5}{},
  cell{7}{6} = {r=4}{},
  cell{12}{1} = {r=3}{},
  cell{15}{1} = {r=5}{},
  cell{15}{6} = {r=5}{},
  vline{2-6} = {1-19}{},
  vline{3-6} = {3-6,8-11,13-14,16-19}{},
  hline{1,20} = {-}{0.08em},
  hline{2} = {-}{},
  hline{3-5,7,11,13-15} = {2-6}{},
  hline{7,15} = {1}{0.03em},
  hline{12} = {1-2}{0.03em},
  hline{12} = {3-6}{},
}
\textbf{RTOS} & \textbf{ID} & \textbf{SRC} & \textbf{ASM} & \textbf{RTOS} & \textbf{SDK} \\
FreeRTOS      & f1          & 88k          & 2.2k         & 105k          & 100k         \\
              & f2          & 22k          & 2.6k         & 13.5k         & 2.16M        \\
              & f3          & 32.4k        & 1k           & 10.3k         & 130k         \\
              & f4          & 657k         & 1.5k         & 209k          & 65k          \\
              & f5          & 656k         & 2k           & 209k          &              \\
Nuttx         & n1          & 429k         & 26k          & 1.7M          & 8k           \\
              & n2          & 428k         & 22k          & 1.6M          &              \\
              & n3          & 429k         & 23k          & 1.65M         &              \\
              & n4          & 429k         & 25k          & 1.7M          &              \\
              & n5          & 310k         & 600          & 1.5M          & 198k         \\
Zephyr        & z1          & 200          & 1k           & 19k           & 0            \\
              & z2          & 5.6k         & 0            & 20k           & 3k           \\
              & z3          & 14.2k        & 0            & 20k           & 2.4k         \\
Threadx       & t1          & 413k         & 52.1k        & 351k          & 335k         \\
              & t2          & 236k         & 52.2k        & 351k          &              \\
              & t3          & 333k         & 51.3k        & 351k          &              \\
              & t4          & 185k         & 51.4k        & 351k          &              \\
              & t5          & 310k         & 52k          & 351k          &              
\end{tblr}
\caption{\highlighted{Breakdown of Source Lines of Code by source files, assembly (inline + standalone)
, RTOS, and SDK (counted once per application). See\mbox{~\tbl{tab:full_dataset_table}} (Appendix) for descriptions of each application.}}
\label{tab:dataset_detail_table}
\end{table}

\Revision{3.3 Split Table 1 to detail per-component SLOC. SDK components counted only once.
}{Done}
\textbf{Dataset:}
\highlighted{\mbox{\tbl{tab:dataset_detail_table}} gives details of our application set with per-component SLOC, selected in 2 steps.}

First, to study Type-2 embedded systems, we ensured diversity at the RTOS level, choosing four popular \acp{RTOS}: FreeRTOS (widely used in resource-constrained systems), Zephyr (modularity and scalability), Nuttx (POSIX-compliant and versatile), and ThreadX (optimized for high-performance real-time applications).

Second, we sampled applications for each RTOS from GitHub, including major, actively maintained projects (e.g., PX4 for drones, Infinitime for smartwatches) and smaller, peripheral-focused ones (e.g., TinyUSB for USB, Nrf\_Pwm for PWM tasks).

\noindent\textbf{Setup:}
We have conducted our experiments on an AMD EPYC 7543P 32 Core Processor with 64 threads and 128 GB of RAM.
In whole program mode, we fuzzed each application for \numperapplicationtime{} hours following suggested best practices \cite{klees2018evaluating}. 
In function-level mode, we fuzzed each target function for \numperfunctiontime{} minutes, after which coverage plateaued for most functions.

\subsection{RQ1: Converting to Linux Applications}
\label{subsec:rq1eval}

\subsubsection{Methodology}

We measure the ability of \systemname{} to successfully convert the 18 embedded applications in our dataset to Linux applications and the amount of manual effort required.
An application is successfully converted if it can be compiled and executed on a Linux operating system without crashing. 
Additionally, as described in \sect{subsubsec:interactivecompilererrorresolution}, \systemname{} relies on human intervention to guide retargeting to desktop ISA. We categorize the required human effort into three categories as follows: 
    (a) Setup (Identifying source files and build/compilation instructions), 
    (b) Addressing errors due to compiler incompatibilities, and
    (c) Handling inline assembly.
We measure the time spent in each category and the impact on the application's source files.

These conversions were performed by the authors, who are graduate students with intermediate expertise in C/C++ but with less experience with embedded codebases.
The conversion time to~\elapp{} depends on familiarity with the embedded codebase, so the reported measurements represent an upper bound; engineers with embedded expertise should require significantly less time.

\begin{figure}[ht]
    \footnotesize
    \centering
    \begin{tikzpicture}
      \begin{axis}
        [
        xmin=0,
        xmax=13,
        xtick={1,2,3,4,5,6,7,8,9,10,11,12},
        xticklabels={,FreeRTOS,,,Nuttx,,,Zephyr,,,Threadx},
        ymin=0,
        ymax=85,
        width=.45\textwidth,
        height=.2\textheight,
        ytick={0,20,40,60,80,100},
        xlabel={RTOSes},
        ylabel={Time (in mins)},
        ]
        \addplot+[black, solid, fill=red!20,
        boxplot prepared={
          lower whisker=20,
          lower quartile=30,
          median=40,
          upper quartile=50,
          upper whisker=60
        },
        boxplot/draw direction=y,
        ] coordinates {};
        \addplot+[black, solid, fill=green!20,
        boxplot prepared={
          lower whisker=5,
          lower quartile=10,
          median=35,
          upper quartile=60,
          upper whisker=80
        },
        boxplot/draw direction=y,
        ] coordinates {};
        \addplot+[black, solid, fill=blue!20,
        boxplot prepared={
          lower whisker=10,
          lower quartile=15,
          median=20,
          upper quartile=25,
          upper whisker=30
        },
        boxplot/draw direction=y,
        ] coordinates {};
        \addplot+[black, solid, fill=red!20,
        boxplot prepared={
          lower whisker=20,
          lower quartile=30,
          median=40,
          upper quartile=50,
          upper whisker=60
        },
        boxplot/draw direction=y,
        ] coordinates {};
        \addplot+[black, solid, fill=green!20,
        boxplot prepared={
          lower whisker=5,
          lower quartile=5,
          median=5,
          upper quartile=5,
          upper whisker=50
        },
        boxplot/draw direction=y,
        ] coordinates {};
        \addplot+[black, solid, fill=blue!20,
        boxplot prepared={
          lower whisker=10,
          lower quartile=15,
          median=20,
          upper quartile=25,
          upper whisker=30
        },
        boxplot/draw direction=y,
        ] coordinates {};
        \addplot+[black, solid, fill=red!20,
        boxplot prepared={
          lower whisker=50,
          lower quartile=60,
          median=60,
          upper quartile=70,
          upper whisker=70
        },
        boxplot/draw direction=y,
        ] coordinates {};
        \addplot+[black, solid, fill=green!20,
        boxplot prepared={
          lower whisker=15,
          lower quartile=20,
          median=30,
          upper quartile=40,
          upper whisker=40
        },
        boxplot/draw direction=y,
        ] coordinates {};
        \addplot+[black, solid, fill=blue!20,
        boxplot prepared={
          lower whisker=15,
          lower quartile=25,
          median=25,
          upper quartile=30,
          upper whisker=35
        },
        boxplot/draw direction=y,
        ] coordinates {};
        \addplot+[black, solid, fill=red!20,
        boxplot prepared={
          lower whisker=40,
          lower quartile=45,
          median=45,
          upper quartile=60,
          upper whisker=60
        },
        boxplot/draw direction=y, 
        ] coordinates {};
        \addplot+[black,solid, fill=green!20,
        boxplot prepared={
          lower whisker=10,
          lower quartile=15,
          median=15,
          upper quartile=20,
          upper whisker=20
        },
        boxplot/draw direction=y,
        ] coordinates {};
        \addplot+[black, solid, fill=blue!20,
        boxplot prepared={
          lower whisker=10,
          lower quartile=15,
          median=15,
          upper quartile=15,
          upper whisker=20
        },
        boxplot/draw direction=y,
        ] coordinates {};
      \end{axis}
    \end{tikzpicture}
    \caption{Comparison of manual effort time (y-axis) across RTOSes (x-axis), categorized into three categories:  (a) \colorbox{red!20}{} - Preconfigurations and source modifications; (b) \colorbox{green!20}{} - Non-automated compiler errors; (c) \colorbox{blue!20}{} - Macro ASM adjustments.}
    \label{fig:eval_times_compare}
\end{figure}
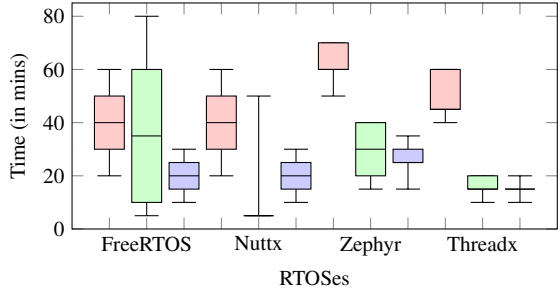
\subsubsection{Results}
Using \systemname{}'s interactive approach, we successfully converted all the applications in our dataset to \elapp{}s.
As shown by the results in \tbl{tab:tool_eval}, we were able to execute and dynamically analyze all 18 applications.

\fig{fig:eval_times_compare} shows the box plot of time spent in each human effort category for applications across different \acp{RTOS}.
In all \acp{RTOS}, setup time (category (a), median 40–60 min) is the largest contributor, mainly for identifying the build setup, dependencies, and toolchains.
These times align with Shen \etal{} \cite{shen2023empirical}, who reported an average of 60 min for embedded build setup.
Note that setup time is independent of \systemname{}.
As shown in \fig{fig:eval_times_compare}, our interactive steps (categories (b) and (c)) contribute minimally to the manual effort.

\tbl{tab:stats_table} highlights SLoC affected for categories (b) and (c).
Manual effort for (b) is relatively low (median 20–40 min) compared to the SLoC modified, demonstrating effective handling of compiler incompatibilities.
For example, NuttX applications required no manual effort for (b) as they used standard C features supported by \clang{}, larger FreeRTOS applications required more effort due to greater SLoC changes.

Despite of a large number of ASM modifications, the amount of manual effort (\ie{} category (c) with a median of 20 - 30 min in \fig{fig:eval_times_compare}) is relatively less, demonstrating the effectiveness of our source code transformations to automatically handle inline assembly.

\summary{RQ1 results demonstrate that \systemname{} can successfully convert embedded applications to \elapp{}s and requires minimal manual effort.}

\begin{table*}[]
\tiny
\centering
\caption{Detailed porting metrics for each application, including type of files modified, lines added or removed, and impact percentages (lines affected over total lines).
The times are summarized in \fig{fig:eval_times_compare}.}
\vspace{0.25cm}
\resizebox{0.8\textwidth}{!}{
\begin{tabular}{c|c|c|c|c|ccc|cc|c}
\toprule
\textbf{AppID}            & \multicolumn{2}{c|}{\textbf{Total Files}}           & \multicolumn{2}{c|}{\textbf{Total Lines}}  & \multicolumn{3}{c|}{\textbf{Files Modified}}                                                                   & \multicolumn{2}{c|}{\textbf{Lines Added/Removed}}                                           & \textbf{Impact \%}             \\ \cmidrule{2-10}
\multicolumn{1}{l|}{\textbf{}} & \multicolumn{1}{c|}{\textbf{App}} & \multicolumn{1}{c|}{\textbf{RTOS}} & \multicolumn{1}{c|}{\textbf{App}} & \multicolumn{1}{c|}{\textbf{RTOS}}  & \multicolumn{1}{c|}{\textbf{Sources}}          & \multicolumn{1}{c|}{\textbf{Headers}}          & \textbf{ASM} & \multicolumn{1}{c|}{\textbf{Sources + Headers}}                              & \textbf{ASM} & \multicolumn{1}{l}{\textbf{}} \\ 
\cmidrule{9-10}

\multicolumn{1}{l|}{\textbf{}} & \multicolumn{1}{c|}{\textbf{}} & \multicolumn{1}{c|}{} & \multicolumn{1}{c|}{\textbf{}} & \multicolumn{1}{c|}{} & \multicolumn{1}{c|}{\textbf{}} &  \multicolumn{1}{c|}{} & \multicolumn{1}{c|}{} & \multicolumn{1}{c|}{\textbf{Category (b)}}                              & \textbf{Category (c)} & \multicolumn{1}{l}{\textbf{}} \\ 
\midrule

\rowcolor[HTML]{EFEFEF}
f1                     & 1.1k &         612                & 184k & 105k              & \multicolumn{1}{c|}{2}                         & \multicolumn{1}{c|}{8}                         & 6            & \multicolumn{1}{c|}{2, -2}                                                   & +505, -964   & 0.51                           \\
f2                      & 5.8k  &           100            & 2.38M  &    13.5k      & \multicolumn{1}{c|}{2} & \multicolumn{1}{c|}{3} & 11           & \multicolumn{1}{c|}{+218, -6} & +410, -1483  & 0.09                           \\

\rowcolor[HTML]{EFEFEF}
f3                      &      201    &     31           & 162.5k    &    10.3k    & \multicolumn{1}{c|}{10} & \multicolumn{1}{c|}{5} & 4           & \multicolumn{1}{c|}{+502, -128} & +539, -953  & 1.23                           \\

f4                   & 232            &       584       & 724k     &   209k    & \multicolumn{1}{c|}{3}                         & \multicolumn{1}{c|}{2}                         & 4            & \multicolumn{1}{c|}{+239, -2}                                                & +539, -1007  & 0.19                      \\

\rowcolor[HTML]{EFEFEF} 
f5                 & 230         & 584  &      723k          &   209k     & \multicolumn{1}{c|}{2} & \multicolumn{1}{c|}{1} & 4            & \multicolumn{1}{c|}{+238, -2}                        & +539, -1007  & 0.19                           \\ \midrule

n1                   & 400             &     14k       & 429k   & 1.7M      & \multicolumn{1}{c|}{0} & \multicolumn{1}{c|}{0} & 4            & \multicolumn{1}{c|}{NIL}                          & +133, -500    & 0.03                           \\

\rowcolor[HTML]{EFEFEF} 
n2                    & 378           &    13.9k       & 428k        &  1.6M   & \multicolumn{1}{c|}{0}                         & \multicolumn{1}{c|}{0}                         & 4            & \multicolumn{1}{c|}{NIL}                                                     & +133, -500   & 0.03                           \\
n3                        & 355            &      13.8k      &   429k   &    1.65M  & \multicolumn{1}{c|}{0} & \multicolumn{1}{c|}{0} & 7            & \multicolumn{1}{c|}{NIL}                             & +147, -537   & 0.03                           \\

\rowcolor[HTML]{EFEFEF} 
n4                         & 400          &       14k       & 429k      &    1.7M   & \multicolumn{1}{c|}{0}                         & \multicolumn{1}{c|}{0}                         & 8            & \multicolumn{1}{c|}{NIL}                                                     & +148, -575   & 0.03                           \\

n5                      & 455           &     12.9k      & 435k      &  1.5M    & \multicolumn{1}{c|}{3} & \multicolumn{1}{c|}{2} & 1            & \multicolumn{1}{c|}{+6, -6}                             & +152, -400   & 0.03                           \\ \midrule

\rowcolor[HTML]{EFEFEF} 
z1                      & 50   &    6.2k       & 200      &  20k   & \multicolumn{1}{c|}{7} & \multicolumn{1}{c|}{3} & 0            & \multicolumn{1}{c|}{+235, -171}     & NIL   & 1.27                           \\

z2     &      203    &   6.2k    & 8.6k  &   20k  & \multicolumn{1}{c|}{7} & \multicolumn{1}{c|}{3} & 0            & \multicolumn{1}{c|}{+220, -180}                             & NIL   & 1.40                           \\

\rowcolor[HTML]{EFEFEF} 
z3                      &      221     &   6.2k   & 16.6k     &   20k   & \multicolumn{1}{c|}{7} & \multicolumn{1}{c|}{3} & 0            & \multicolumn{1}{c|}{+220, -180}                             & NIL   & 1.09                           \\ \midrule

t1                      & 4.8k       &       1.3k       &    748k      &   351k   & \multicolumn{1}{c|}{1}                         & \multicolumn{1}{c|}{0}                         &     2            & \multicolumn{1}{c|}{+3, 0}                                                   & +568, -884   & 0.14                           \\
\rowcolor[HTML]{EFEFEF} 
t2                     & 3.8k           &   1.3k            & 571k       &   351k   & \multicolumn{1}{c|}{0} & \multicolumn{1}{c|}{0} & 2            & \multicolumn{1}{c|}{NIL}                             & +569, -879   & 0.16                           \\
t3              & 3.5k          &      1.3k          &  668k     &      351k & \multicolumn{1}{c|}{1}                         & \multicolumn{1}{c|}{0}                         & 2            & \multicolumn{1}{c|}{+4, 0}                                                   & +568, -884   & 0.14                           \\
\rowcolor[HTML]{EFEFEF} 
t4                       & 3.3k           &      1.2k       & 520k        &  351k  & \multicolumn{1}{c|}{0} & \multicolumn{1}{c|}{0} & 1            & \multicolumn{1}{c|}{NIL}                             & +23, -167    & 0.02                           \\
t5              & 4.3k           &    1.2k          & 645k       &     351k & \multicolumn{1}{c|}{1}                         & \multicolumn{1}{c|}{0}                         & 2            & \multicolumn{1}{c|}{+3, 0}                                                   & +568, -884   & 0.14                           \\ 
\bottomrule
\end{tabular}
}
\label{tab:stats_table}
\end{table*}

\vspace{-6pt}
\subsection{RQ2: Peripheral Handling}
\label{subsubsec:peripheraleval}

\subsubsection{Methodology}

We assess the effectiveness of our constant address analysis in two ways.
First, we validate discovered address ranges by checking for overlaps with the \elapp{}'s actual memory map, ensuring MMIO ranges remain distinct.
This validation leverages standardized memory boundaries documented in CMSIS-SVD files, which define peripheral registers and their address mappings.

Second, we corroborate the results of \systemname{}'s constant address analysis by comparing the discovered address ranges against those specified in CMSIS-SVD files~\cite{martin2016designer}.
Discrepancies are manually investigated through random sampling to understand gaps in identification.
Both methods are necessary to ensure accuracy and to identify limitations of SVD-based documentation versus our constant address analysis approach.

\subsubsection{Results}

The \tbl{tab:mmio_table} shows the number of MMIO address ranges found across different applications.
Upon investigation, we found that none of these address ranges conflict with the memory map of the corresponding \elapp{}.
Hence, instrumenting reads from these addresses should not affect \elapp{}'s execution.

When we compared the discovered address ranges with those in CMSIS-SVD files \cite{martin2016designer}, we found that over 50\% of our address ranges are missing in CMSIS-SVD files (last column of \tbl{tab:mmio_table}).
Further analysis revealed that the missing address ranges represented valid MMIO addresses in the source code and corresponded with those used by valid core peripherals~\cite{SVDMissing}.
\lst{lst:mmio_extra} (Appendix) shows MMIO address ranges used in the codebase but missing from the corresponding peripheral’s SVD file.

\summary{RQ2 results show that our constant address analysis is effective at finding MMIO address ranges and provides more complete results than the commonly used approach of analyzing SVD files.}

\subsection{RQ3: Testing~\elapp{}s}
\label{subsec:testingelapp}

\subsubsection{Methodology}

In this RQ, we assess the effectiveness of the converted \elapp{}s in supporting different fuzzing modes: Whole Program Fuzzing with MMIO instrumentation (M1), Whole Program Fuzzing with MMIO + Weakening State-Dependent Conditions (M2), and Function-Level fuzzing (M3) incorporating all optimizations from (M1) and (M2). We measure and report the code coverage (in terms of unique basic blocks covered) and the number of unique bugs detected through each mode, following crash triaging and manual confirmation according to our threat model.
For whole-program fuzzing, we identified the MCU firmware ELF entrypoint and ensured the \elapp{}'s entrypoint matched it (ignoring assembly-based entrypoints). This was necessary to initialize global structures for peripheral handling and avoid \inlinecode{NULL}-ptr dereferences in the \elapp{}.
To identify candidate functions for function-level fuzzing, we first filtered functions that take pointer arguments without a specified size. Next, we manually verified (5 minutes per function) whether these functions performed any interesting operations, such as pointer arithmetic or explicit casts, which are common in risky programming idioms. Previous work \cite{du2019leopard} indicates that these characteristics are strong indicators of potentially buggy functions. Depending on the target, this process typically leaves us with roughly 100-150 functions per application for further fuzzing.


\subsubsection{Results}
\begin{table}[t]
\footnotesize
\renewcommand{\arraystretch}{0.9}
\setlength{\tabcolsep}{3pt}
\centering
\begin{tabular}{ccc}
\toprule
\textbf{AppID} & \textbf{Detected MMIOs} & \textbf{In SVD (\% of Detected)}  \\
\midrule
\rowcolor[HTML]{EFEFEF} f1 & 45 & 11 (24.44) \\
f2 & 33 & 16 (48.48) \\
\rowcolor[HTML]{EFEFEF} f3 & 35 & 18 (51.43) \\
f4 & 15 & 11 (73.33)  \\ 
\rowcolor[HTML]{EFEFEF} f5 & 15 & 11 (73.33)  \\ 
\midrule
n1 & 8 & 4 (50.0) \\
\rowcolor[HTML]{EFEFEF} n2 & 10 & 5 (50.0)  \\
n3 & 9 & 4 (44.44)  \\
\rowcolor[HTML]{EFEFEF} n4 & 9 & 4 (44.44)  \\
n5 & 60 & 9 (15.0)  \\ 
\midrule
\rowcolor[HTML]{EFEFEF} z1 & 10 & 4 (40.0)  \\
z2 & 10 & 5 (50.0) \\ 
\rowcolor[HTML]{EFEFEF} z3 & 54 & 25 (46.3) \\ 
\midrule
t1 & 16 & 6 (37.5) \\
\rowcolor[HTML]{EFEFEF} t2 & 16 & 6 (37.5) \\
t3 & 3 & 1 (33.33) \\
\rowcolor[HTML]{EFEFEF} t4 & 16 & 6 (37.5) \\ 
t5 & 16 & 6 (37.5) \\
\bottomrule
\end{tabular}
\caption{MMIO detection analysis highlights potential undocumented peripherals in SVD files. SVD Detection shows documented MMIOs, while Potential MMIOs indicates detected MMIOs that may represent undocumented peripherals.}
\label{tab:mmio_table}
\end{table}
\tbl{tab:tool_eval} shows the code coverage and bug detection results when conducting whole program ($M2$) and function-level ($M3$) fuzzing.
Using \systemname{}, we conducted whole program and function-level fuzzing on 15 applications each.
We manually created the memory layout for z1 as a demonstration but did not perform whole-program fuzzing for z2, z3, and t3 due to their layout-dependent code, which is not automated (\sect{sec:limitationsfuturework}).
We did not perform function-level fuzzing on f1, f2, and n5 as they were written in C++, and our current implementation of function-level fuzzing does not support C++ objects.

\textit{\textbf{Code Coverage:}}
In whole program fuzzing, we triggered a considerable number of reachable functions (\ie{} those that can be reached through \inlinecode{main}) in each \elapp{}.
\fig{fig:app_coverage} shows the percentage of triggered functions, with over 70\% triggered on average, except for f2, f3, and z1.
The \ac{CDF} in \fig{fig:app_coverage} illustrates function coverage, where each point $(x, y)$ indicates that $x\%$ of triggered functions have $y\%$ or less code coverage.
The consistent slope across \elapp{}s confirms that \systemname{} enables effective testing with reasonable coverage.
For example, in FreeRTOS \elapp{}s, 40\% of triggered functions achieve 40\% or more code coverage.
\tbl{tab:tool_eval} shows absolute coverage, with function-level fuzzing providing $\sim$10x more coverage than whole-program fuzzing, as it targets individual functions. \\ \vspace{0.1cm}
\textit{\textbf{Bug Detection:}}
\tbl{tab:tool_eval} also shows the bugs detected by each approach.
Overall, as expected, function level fuzzing ($M3$) identified \nummorebugsfunctionlevel{} additional unique bugs.
This is due to its ability to directly exercise risky functions.
As shown in \lst{lst:motivation_bug_new}, function-level fuzzing ($M3$) uncovered an out-of-bounds access in a deep function that whole-program fuzzing ($M2$) missed, as the function was never triggered.

As shown in the last row of \tbl{tab:tool_eval}, although total bugs are large (\eg{} \numfunctionleveltotalbugs{}), the number of unique bugs is small (\eg{} \numfunctionleveluniquebugs{}). This is because the same bugs (those in RTOS functions) could be present multiple in \elapp{}s.
More details can be found in \sect{apdx:bug_detection} (Appendix).
\Revision{3.1 Provide detailed descriptions of the bugs that your technique found and developer responses.}{Table 8 in appendix gives all details.}
\highlighted{\mbox{\tbl{tab:bug_details}} (Appendix) summarizes the bug types, affected applications, bug descriptions, and developer responses.}
The \tbl{tab:bug_table} (Appendix) shows a detailed split of bugs and unique bugs.
The \tbl{tab:bug_types} (Appendix) provides the categorization of unique bugs.
We found several memory corruption bugs in addition to robustness bugs, such as Divide by zero (\lst{lst:mmio_instrumentation}).

\summary{RQ3 results demonstrate that \systemname{} facilitates whole-program and function-level fuzzing, leading to high code coverage and bug detection.}
\input{figures/app_coverage}

\subsection{RQ4: Ablation Study}
\label{subsec:ablationstudy}

\subsubsection{Methodology}
This RQ measures the contributions of our peripheral handling (\sect{subsubsec:periphelmodeling}) and condition weakening (\sect{subsubsec:dynamicanalysisassistance}) instrumentation in facilitating dynamic analysis. We disabled each of these instrumentations and report their impact on whole-program fuzzing. While function-level fuzzing performed better, whole-program fuzzing compensated for the limitations of function-level fuzzing on C++ applications and provided insights into how effectively a \elapp{} can be tested as a standalone application.

\subsubsection{Results}

\textit{\textbf{Peripheral handling instrumentations:}}
When the instrumentations on MMIO accesses are disabled, we observed that all \elapp{}s crash immediately after they are started.
As mentioned before, \elapp{}s are fuzzed as regular Linux applications, in which MMIO addresses may not be mapped; consequently, any MMIO accesses will result in invalid memory access and segfault.
This shows that \emph{our peripheral handling instrumentation is necessary for testing \elapp{}s}. \\ [0.5em]
\textit{\textbf{Condition weakening instrumentation:}}
When the instrumentations that weaken state-dependent conditions are disabled, we observe a remarkable drop in the number of covered basic blocks.
The \textbf{M1} and \textbf{M2} columns in \tbl{tab:tool_eval} shows the number of covered basic blocks and bugs found when conducting whole-program fuzzing without and with this instrumentation.
On average, we see an improvement of $\sim$2x in the number of basic blocks covered with M2 over M1.
All bugs found by M1 were also detected by M2, with the addition of \nummorebugsmmiocond{} more bugs.
These results show that embedded applications greatly depend on the peripheral state for their execution, and ignoring them results in ineffective testing.
\summary{RQ4 results show that our instrumentation-based techniques significantly improve the effectiveness of testing.}

\subsection{RQ5: Comparative Evaluation}
\label{subsec:comparativeeval}

\subsubsection{Methodology}
In this RQ, we compare the code coverage and bug detection results of~\systemname{} with results from other recent dynamic analysis techniques that target embedded applications.
We selected baselines that follow the~\systemname{}'s philosophy of being usable on applications without requiring low-level understanding of the application's internal implementation.
This led to three baselines: P2IM \cite{feng2020p2im}, Fuzzware ($Fw$) \cite{fuzzware} and MultiFuzz ($Mf$) \cite{multifuzz}, 
and the exclusion of three others: PMCU \cite{li2021library}, Halucinator~\cite{clements_halucinator_2020} and SAFIREFUZZ~\cite{seidel2023forming}.
Notably, PMCU required a custom RTOS configuration for each application based on the application's internal implementation. 
Halucinator and SAFIREFUZZ require creating handlers for each peripheral the application interacted with.

We were unable to set up P2IM, despite following their instructions and attempting to contact the authors.
Additionally, we encountered challenges setting up fuzzing for certain applications (\eg{} f3, z3) using Fuzzware and Multifuzz, primarily due to inaccurate memory modeling, resulting in applications crashing with unsupported or invalid instructions.
Consequently, we evaluated Fuzzware and Multifuzz on the remaining original, non-converted \elapp{}s.
\vspace{-0.5cm}
\subsubsection{Results}
\tbl{tab:tool_eval} shows the results of fuzzing each \elapp{} with the selected baselines. We found that, on average, \systemname{} configurations (M2/M3) detected 21 bugs, while Multifuzz (\textit{Mf}) and Fuzzware (\textit{Fw}) detected 1 and 3 bugs, respectively.
From a code coverage perspective, \textit{Mf} outperformed \textit{Fw} for most applications. However, \systemname{} configurations (M2/M3) outperformed \textit{Mf} for most applications except for n1 and z2.
This was primarily due to \textit{Mf}'s ability to trigger nested interrupts, which led to higher coverage.
In contrast, \systemname{} uses a simpler interrupt handling approach (\sect{subsubsec:periphelmodeling}) and does not support nested interrupts.
We also observed that \textit{Fw} occasionally reported false positives, such as crashes in z1, caused by incorrect interrupt handling.
%
From the bug detection perspective, \systemname{} is even more effective by detecting \numbugs{} bugs, with Fuzzware and MultiFuzz detecting only 1 and 3 respectively.
Furthermore, all bugs found by Fuzzware and MultiFuzz are also found by \systemname{}.

\summary{RQ5 results indicate that \systemname{} has better bug-finding ability than existing techniques.}

\subsection{RQ6: False Positive Analysis}
\label{subsec:falsepositives}

\subsubsection{Methodology}

This RQ aims to analyze false positives that arise due to our low-fidelity approximations, how they compare with existing works, and provide remediation for each type of false positive.
Our analysis covers the \systemname{} components that introduce approximations, as these are the sources of false positives.

\subsubsection{Results}
\highlighted{\mbox{\tbl{tab:fp_low_fidelity}} summarizes the number and type of false positives encountered across various applications.}


\begin{table}[]
\tiny
\centering
\resizebox{0.3\textwidth}{!}{
\begin{tabular}{cccc}
\toprule
\textbf{AppID} & \textbf{Inline ASM} & \textbf{Interrupt Misfiring} & \textbf{Board Layout} \\ [0.3em] \toprule
\rowcolor[HTML]{EFEFEF} 
f1                    & 1                                      & 1                     & 0                                                         \\
f2                    & 1                                    & 3                     & 0                                                        \\ 
z2                    & 0                                   & 0                     & 1                                                        \\
\rowcolor[HTML]{EFEFEF} 
z3                    & 0                                    & 0                     & 1                                                        \\ 
t2                    & 0                                    & 1                     & 1                                                        \\ \midrule
\rowcolor[HTML]{EFEFEF} \textbf{Total}        & \textbf{2}          & \textbf{5}            & \textbf{3}                             \\ \bottomrule
\end{tabular}
}
\caption{\highlighted{False positives due to \systemname{} approximations.}}
\label{tab:fp_low_fidelity}
\end{table}



\highlighted{\textit{\textbf{POSIX Swap (\mbox{\sect{subsubsec:handlingexecsemantics}}):}} \systemname{} replaces the board-specific RTOS layer with a POSIX-compatible one. While this can introduce false positives due to kernel misconfiguration, we observed none in our case. For example, incorrect task priorities could affect behavior, but we mitigated this by incorporating all relevant application kernel configurations.}

\highlighted{\textit{\textbf{Inline ASM (\mbox{\sect{subsubsec:interactivecompilererrorresolution}}):}} \systemname{} removes all inline assembly and approximates expected values at runtime using random values of the same type 
This led to two false positives across all applications.
For instance, \mbox{~\lst{lst:inlineASM}} (Appendix) shows how inline assembly was used for initialization, which we identified through debugger-inspected halts.
We resolved this by manually patching the instruction to return the expected value.}

\highlighted{\textit{\textbf{Symbol Modifications (\mbox{\sect{subsubsec:oppsymbolresolution}}):}} Our symbol modification strategy iteratively resolves linker errors and could, in principle, introduce false positives such as from incorrect renaming of indirect function calls.
However, we observed none. 
}

\highlighted{\textit{\textbf{Interrupt Misfiring  (\mbox{\sect{subsubsec:periphelmodeling}})}}: \systemname{} attempts to trigger all interrupts from the \textit{isr\_vector\_table}, which can cause crashes if global structures containing callback routines are uninitialized.
For example, \mbox{\lst{lst:isr_fp}} (Appendix) shows a misfired interrupt caused by this dependency. We remediate using lightweight static analysis to trigger interrupts accurately.}

\highlighted{\textit{\textbf{Board Layout (\mbox{\sect{sec:limitationsfuturework}}):}} Board-specific layout-dependent code is a limitation of our work, preventing analysis of two Zephyr and one ThreadX applications.
An example of this issue is shown in \mbox{\lst{lst:board_layout}}.
We manually constructed the layout for one Zephyr application for our evaluation.}

\highlighted{The existing tools used in our Comparative Evaluation (\mbox{\sect{subsec:comparativeeval}}) also suffer from false positives due to misfired interrupts and emulation issues.
In contrast to \systemname{}, these tools had a greater number of false positive crashes, as indicated by the large numbers along the crash column of \mbox{\tbl{tab:tool_eval}}.
Furthermore, triaging these crashes (especially in the case of Fuzzware (\textit{Fw})) is non-trivial, and has also been acknowledged by recent work\mbox{~\cite{FirmRCA}}.
\systemname{} adopts an approximate but principled approach, enabling us to easily identify false positives and resolve them.}

\summary{\highlighted{RQ6 results indicate that the lower-fidelity execution model used by \systemname{} does not lead to a large number of false positives, as evidenced by our comparative evaluation.}}

\section{Limitations and Future Work}
\label{sec:limitationsfuturework}
We recognize the following limitations of \systemname{} and plan to handle them as part of our future work.
\begin{itemize}[leftmargin=*, noitemsep]

\item\textbf{Dependency on \ac{LPL}:}
Our approach depends on the existence of \ac{LPL} for \acp{RTOS}.
As shown in \apdx{subsubsec:prevalenceofnpl}, most \acp{RTOS} already have \ac{LPL}, we argue that it is fairly easy to create \ac{LPL} based on existing implementations.
\item\textbf{Incomplete ISR coverage:}
Our approach identifies \acp{ISR} via RTOS-specific pattern matching, which worked reliably in our experiments.
However, we skip ISRs that depend on global state, leading to some coverage gaps.
Recent works like AIM~\cite{aim} improves ISR identification and invocation, and we plan to incorporate such techniques into \systemname{} in future work.
\vspace{0.1cm}
\item\textbf{Unable to handle layout-specific code:}
We found cases where embedded applications rely on specific memory layouts, hindering our efforts in further analysis. \lst{lst:board_layout} (Appendix) shows an example from a Zephyr \ac{RTOS} application.
As future work, we plan to automatically detect and refactor such code idioms.
\end{itemize}

\section{Related Work}
\label{sec:relatedwork}

Dynamic analysis techniques, especially automated random testing or fuzzing~\cite{FuzzingSurvey, godefroid2020fuzzing}, are demonstrated to be effective at vulnerability detection.
\emph{Rehosting} is a necessary requirement for scalable dynamic analysis.
This process is relatively easy for Type-1 systems~\cite{ChenWBE16, FirmGuide}, \ie{} those based on standard OSes such as Embedded Linux.
Consequently, several techniques~\cite{fasano2021sok} exist for rehosting Type-1 systems.
But, these cannot be applied to Type-2 systems because of lack of well-defined OS interface and tight coupling with hardware~\cite{fasano2021sok}.

One of the most important challenges of Rehosting Type-2 systems is the capability to handle peripheral interactions.
Existing techniques to handle this can be categorized at a high level into hardware-in-the-loop~\cite{kammerstetter_prospect_2014} or software model~\cite{spensky2021conware} based approaches.
The hardware-in-the-loop approaches~\cite{muench_avatar_2018,li_femu_2010,ruge_frankenstein_2020,gui_firmcorn_2020,corteggiani_inception_nodate, kammerstetter_embedded_2016, kammerstetter_prospect_2014, koscher_surrogates_2015} achieve the highest level of fidelity and less manual effort. Given the diversity of hardware platforms, these techniques are hard to scale.

The software-only approaches~\cite{feng2020p2im, uEmu, fuzzware, Kim2020FirmAETL, Srivastava2019FirmFuzzAI} provide low-fidelity execution unless there are precise peripheral models.
Automated peripheral modeling techniques~\cite{clements_halucinator_2020,feng2020p2im,spensky2021conware, gustafson2019toward} are specific to certain peripherals and hard to generalize.
Some techniques~\cite{fuzzware, corteggiani_inception_nodate} use symbolic execution~\cite{noauthor_klee_nodate} to create peripheral models. As shown by the recent systematization work~\cite{fasano2021sok}, these techniques are hard to extend for different peripherals and depend on the existence of emulators ~\cite{QEMU, magnusson_simics_2002} of the corresponding~\ac{ISA}.
On the other hand, works such as {\sc MetaEmu}~\cite{chen2022metaemu} attempt to rehost firmware in an architecture-agnostic way by lifting firmware code to an Intermediate Representation as directed by Ghidra's Language Specifications \cite{rohleder2019hands} enabling multi-target analysis. However, these techniques struggle with manual efforts required for specification creation, peripheral modeling, and limited support for specialized automotive protocols.

One of the most closely related works is by Li\mbox{~\etal{}\cite{li2021library}}, who rehost \mbox{\ac{MCU}} libraries for testing on Linux by implementing a portable MCU (PMCU) using the POSIX interface and abstracting hardware functions.
However, their method relies on hand-written abstractions for specific libraries and does not handle unknown or undocumented peripherals, nor does it scale well across diverse firmware binaries.
\systemname{} side-steps the problem of precise peripheral emulation by using \ac{NPL}, which relaxes the requirement of precise peripheral models without affecting the execution of the target embedded system.
\Revision{2. Novelty issue: further clarify the contributions of their approach, especially the novelty part.}{Substantiating novelty claims.}
\highlighted{Unlike prior works that require accurate peripheral models or emulation for specific hardware targets, our approach generalizes across firmware by focusing on what is sufficient to trigger bugs, rather than replicating exact hardware behavior.
As a result, dynamic analysis can be applied to embedded code in a more generalizable manner.}
\section{Conclusion}
We propose \systemname{}, a novel approach to rehosting embedded applications as Linux applications by providing solutions to the associated challenges of retargeting to \targetarch{}, preserving the execution semantics, and handling the peripheral interactions.
We evaluated \systemname{} on \numapp{} embedded applications across \numrtos{} \acp{RTOS} and found \numbugs{} previously unknown bugs, most of which are confirmed and fixed by the corresponding developers.
Our comparative evaluation shows that \systemname{} outperforms existing state-of-the-art techniques in testing embedded applications.

\newpage
\textit{The call for papers states that an extra page is allotted to discuss ethics considerations and compliance with the open science policy. This page contains that content.}
\section{Ethics Considerations}
This section describes the ethical considerations involved in designing, implementing, and evaluating our proposed system.
We identify two relevant classes of stakeholders in this work.
\begin{itemize}
    \item Maintainers of the \acp{RTOS} and embedded applications on which we evaluated.
    \item Users of the \acp{RTOS} and embedded applications on which we evaluated.
\end{itemize}

These stakeholders share in the following risks and benefits:

\begin{itemize}
\item \textit{Risks and benefits from discovered bugs:}
To evaluate \systemname{}, we applied it to 18 applications and uncovered 21 distinct defects in both the applications and the underlying \acp{RTOS}.
Under our threat model, these defects could lead to malfunctions or crashes in the affected software.
We reported all identified issues to the respective vendors and provided GitHub patches to facilitate their resolution.
In our assessment, we did not perceive these defects posed a security threat, so we followed the projects' standard defect disclosure processes (public issue reports) rather than their security vulnerability disclosure processes.
Most of the \textit{maintainers} acknowledged the reported bugs, and our proposed patches were accepted and merged.
In consequence, any \textit{users} who  do not update to the latest versions may face a security risk if attackers can exploit these bugs.
\item \textit{Risks and benefits from the existence of \systemname{}:}
Beyond our evaluation, \systemname{} will be an open-source tool (see~\S\ref{sec:openscience}). Like any defect discovery tool, \systemname{} can be used by maintainers and users in good faith, or abused by malicious actors seeking vulnerabilities to exploit.
\end{itemize}

These stakeholders and risks/benefits are common to all defect detection systems, \eg the fuzzing literature.
The cybersecurity research community understands this risk-benefit tradeoff to fall within the scope of ethical practice for cybersecurity research.

To further mitigate any potential risks, we submitted patches with each bug report, to help the developers fix the vulnerabilities promptly.
None of the affected vendors assessed the reported issues as having significant security implications --- our patches were typically merged but no CVEs were issued.

\section{Open Science} 
\label{sec:openscience}

We have released a raw development version of \systemname{} along with the dataset and necessary documentation at : \toolurl{}.


\section{Acknowledgments}
We would like to thank the reviewers and our shepherd for their valuable comments and inputs to improve our paper. This research was supported by Rolls-Royce and the National Science Foundation (NSF) under Grant CNS-2340548. Any opinions, findings, conclusions, or recommendations expressed in this material are those of the author(s) and do not necessarily reflect the views of Rolls-Royce and NSF.


\bibliographystyle{plain}
\bibliography{paper}


\appendix
\section{Outline of Appendices}

\begin{itemize}
    \item Portable Layer and Implementation details – different phases of \systemname{} (\ref{apdx:appendix_b} Appendix)
    \item Dataset details and bugs found by \systemname{} (\ref{apdx:appendix_c} Appendix)
    \item Effects of removed inline assembly, instrumentation to handle interrupts, memory layout execution. (\ref{apdx:appendix_d} Appendix)
    \item Details of Compiler toolchain differences. (\ref{apdx:appendix_e} Appendix)
    \item Missing SVD Files and comparison against state of the art (\ref{apdx:appendix_f} Appendix)
\end{itemize}

\section{Appendix}
\label{apdx:appendix_b}

\subsubsection{Examples of Continuous Scale Gradations}
\label{apdx:continuousfidelity}
\begin{itemize}
\item \textbf{S Fidelity:} Variable-length arrays are supported by embedded GCC-based compilers but lack support in clang.
\item \textbf{A Fidelity:} An emulator that implements all base instructions of ARMv7 but fails to emulate \textbf{SIMD} instructions or other processor extensions.
\item \textbf{P Fidelity: } An emulator that handles GPIO and UART correctly but approximates behavior for DMA.
\item \textbf{C Fidelity:} A simulation environment allows basic interrupt handling but does not handle nested interrupts.
\end{itemize}

\subsubsection{\acf{NPL}}
\label{subsubsec:npldetails}



To improve testability and aid embedded firmware development, many RTOSes also provide ports for various host operating systems such as Linux and Windows.
We refer to these ports as the \emph{\acf{NPL}} and this includes the \emph{\acf{LPL}} and the \emph{\acf{WPL}}. These native ports allow embedded applications built on these RTOSes to be run on respective desktop operating systems as native applications. 
Native ports use host-provided implementation to simulate various embedded functionalities. For example, the \acf{LPL} of the FreeRTOS~\cite{FreeRTOS} and Zephyr~\cite{ZephyrRTOS} operating systems use Linux \emph{pthreads} to simulate tasks, \emph{signals} to simulate interrupts, and \emph{timers} to simulate clocks in the application.
Some native ports contain only implementations for core RTOS functions such as task management, task context switching, interrupts, timers, and counters while other ports provide simulated implementations of various non-essential peripherals as well such as communication buses (SPI, I2C) and IO devices (USB, buttons, keyboard, LED, etc.).

\subsubsection{Prevalence of \ac{NPL}}
\label{subsubsec:prevalenceofnpl}
A review of 23 actively maintained open-source RTOSes (as listed on the OSRTOS page~\cite{osrtoses}) shows that 14 (60\%) RTOSes provide native ports. 
NPLs provided by 10 RTOSes only contain implementations of core RTOS functions, while 4 RTOSes also provide simulated implementations of non-essential peripherals also.
Our review of the implementations of the available \acf{NPL} also shows that while each native port is designed according to the architecture of their underlying RTOS, they have similar implementations as they rely on similar Linux or Windows features.
\begin{table*}
\scriptsize
\centering
\renewcommand{\arraystretch}{1.5}
\begin{tabular}{c|l|c|p{0.5\linewidth}|c}
\toprule
\textbf{RTOS} & \multicolumn{1}{c|}{\textbf{Application}} & \multicolumn{1}{c|}{\textbf{ID}} & \multicolumn{1}{c|}{\textbf{Description}} & \textbf{SLOC} \\ \midrule
\multirow{5}{*}{FreeRTOS} 
& \cellcolor[HTML]{EFEFEF}FlipperZero & \cellcolor[HTML]{EFEFEF}f1 & \cellcolor[HTML]{EFEFEF}Open source multi-tool device for researching and pentesting radio protocols, access control systems, hardware, and more. & \cellcolor[HTML]{EFEFEF}289k \\
& Infinitime & f2 & Firmware for the PineTime smartwatch & 2.39M \\
& \cellcolor[HTML]{EFEFEF}SmartSpeaker & \cellcolor[HTML]{EFEFEF}f3 & \cellcolor[HTML]{EFEFEF}Smart speaker based on cloud speech recognition running on FreeRTOS & \cellcolor[HTML]{EFEFEF}172.8k \\
& cdc\_msc\_freertos (TinyUSB) & f4 & Example application to trigger a communication device class (cdc) task of tinyusb & 933k \\
& \cellcolor[HTML]{EFEFEF}hid composite freertos (TinyUSB) & \cellcolor[HTML]{EFEFEF}f5 & \cellcolor[HTML]{EFEFEF}Open-source cross-platform USB Host/Device stack for embedded systems. Example app to trigger tinyusb hid (Human Interface Device) task & \cellcolor[HTML]{EFEFEF}932k \\ \midrule

\multirow{5}{*}{Nuttx}
& nrf52840-dk pwm Application & n1 & Demonstrates basic PWM support for the nRF52840-DK board. & 2.1M \\
& \cellcolor[HTML]{EFEFEF}nrf52-feather i2c Application & \cellcolor[HTML]{EFEFEF}n2 & \cellcolor[HTML]{EFEFEF}Demonstrates basic I2C support for the nRF52-Feather board. & \cellcolor[HTML]{EFEFEF}2.02M \\
& nsh (Nuttx-apps) & n3 & Ships the entire nuttx kernel as a busybox application with several Unix-like utilities that can be flashed to firmware to access nuttx features in a shell. & 2.07M \\
& \cellcolor[HTML]{EFEFEF}posix spawn (Nuttx-apps) & \cellcolor[HTML]{EFEFEF}n4 & \cellcolor[HTML]{EFEFEF}Demonstrates how to use the posix\_spawn function to create a new process with more control over attributes compared to fork. & \cellcolor[HTML]{EFEFEF}2.1M \\
& PX4-Autopilot & n5 & PX4 flight control solution for drones running which has support for nuttx kernel & 1.9M \\ \midrule

\multirow{3}{*}{Zephyr}
& Zephyr Blinky & z1 & Demonstrates basic GPIO control and the core Zephyr kernel task creation. & 20.2k \\
& \cellcolor[HTML]{EFEFEF}zmk & \cellcolor[HTML]{EFEFEF}z2 & \cellcolor[HTML]{EFEFEF}Zephyr Mechanical Keyboard (ZMK) Firmware. & \cellcolor[HTML]{EFEFEF}28.6k \\
& ZSWatch & z3 & Open Source Zephyr-based Smartwatch firmware. & 36.6k \\ \midrule

\multirow{5}{*}{Threadx} 
& Tx\_FreeRTOS\_Wrapper & t1 & Demonstrates how to develop an application using the FreeRTOS adaptation layer for ThreadX. & 1.09M \\
& \cellcolor[HTML]{EFEFEF}Tx\_LowPower & \cellcolor[HTML]{EFEFEF}t2 & \cellcolor[HTML]{EFEFEF}Demonstrates how to develop an application using the ThreadX low power APIs when coupled with STM32F4xx low power profiles. & \cellcolor[HTML]{EFEFEF}922k \\
& Tx\_Module & t3 & Demonstrates how to load, start, and unload modules and use ThreadX memory protection via the MPU. & 1.01M \\
& \cellcolor[HTML]{EFEFEF}Tx\_Thread\_Creation & \cellcolor[HTML]{EFEFEF}t4 & \cellcolor[HTML]{EFEFEF}Demonstrates how to create/destroy multiple threads using Azure RTOS ThreadX APIs, including preemption thresholds and priority changes on-the-fly. & \cellcolor[HTML]{EFEFEF}871k \\
& Tx\_Thread\_MsgQueue & t5 & Demonstrates how to send/receive messages between threads using ThreadX message queue APIs with event chaining features. & 996k \\
\bottomrule
\end{tabular}
\vspace{0.25cm}
\caption{Dataset involving 18 applications across four prevalent RTOSes with application IDs added for clarity. SLOC represents combined sources from both the application and the underlying RTOS.}
\label{tab:full_dataset_table}
\end{table*}
\begin{table*}
\scriptsize
\centering
\renewcommand{\arraystretch}{1.5}
\begin{tabular}{c|c|p{0.5\linewidth}|c}
\toprule
\rowcolor[HTML]{FFFFFF} 
\textbf{App ID} & \textbf{Bug}   & \multicolumn{1}{c}{\cellcolor[HTML]{FFFFFF}\textbf{Description}}    & \textbf{Status of Bug} \\ \midrule
\rowcolor[HTML]{EFEFEF} 
f2              & Assert Failure & Inconsistent use of configASSERT FreeRTOS Kernel                                                                                                                                       & \textcolor{green}{\faUserCheck}          \\
\rowcolor[HTML]{FFFFFF} 
f2              & Assert Failure & Assert Failure in ble\_event                                                                                                                                                           & \textcolor{orange}{\faUserClock}           \\
\rowcolor[HTML]{EFEFEF} 
f2              & Build Related  & Conflict of min and max from stl\_algo.h in HeartRateService.h                                                                                                                         & \textcolor{green}{\faUserCheck}          \\
\rowcolor[HTML]{FFFFFF} 
f3              & Div By Zero    & FPE in RCC\_GetClocksFreq due to missing MMIO Checks                                                                                                                                   & \textcolor{green}{\faUserCheck}          \\
\rowcolor[HTML]{EFEFEF} 
f3              & Null Deref     & Null Deref in ucFlash\_Write                                                                                                                                                           & \textcolor{orange}{\faUserClock}           \\
\rowcolor[HTML]{FFFFFF} 
f3              & OOB  Write     & Potential undefined behavior on overlapping copy in mem\_cpy                                                                                                                           & \textcolor{orange}{\faUserClock}           \\
\rowcolor[HTML]{EFEFEF} 
f4              & OOB Write      & Potential OOB memcpy in tud\_msc\_read10\_cb                                                                                                                                           & \textcolor{green}{\faUserCheck}          \\
\rowcolor[HTML]{FFFFFF} 
f4              & DoS            & Infinite Loop in port\_event\_handle due to missing MMIO Checks                                                                                                                        & \textcolor{red}{\faUserTimes}               \\
\rowcolor[HTML]{EFEFEF} 
f5              & OOB Read       & Potential OOB Read \& Null Deref due to missing MMIO Checks in board\_get\_unique\_id                                                                                                  & \textcolor{orange}{\faUserClock}           \\ \midrule
\rowcolor[HTML]{FFFFFF} 
n1              & Build Related  & Buggy handling of unsigned long in vsprintf\_internal                                                                                                                                  & \textcolor{green}{\faUserCheck}          \\
\rowcolor[HTML]{EFEFEF} 
n3              & Build Related  & Compilation failure due to improper handling of CAN utils lely-core package                                                                                                            & \textcolor{green}{\faUserCheck}          \\
\rowcolor[HTML]{FFFFFF} 
n4              & OOB Write      & Undefined behaviour on partial overlapping copy in sim\_copyfullstate                                                                                                                  & \textcolor{green}{\faUserCheck}          \\
\rowcolor[HTML]{EFEFEF} 
n5              & DoS            & Infinite Loop in up\_enable\_dcache due to invalid MMIOs                                                                                                                               & \textcolor{orange}{\faUserClock}           \\ \midrule
\rowcolor[HTML]{FFFFFF} 
z1              & OOB Write      & Stack Overflow in buf\_char\_out if CONFIG\_PRINTK\_BUFFER\_SIZE is 0                                                                                                                  & \textcolor{blue}{\faUserPlus}                    \\
\rowcolor[HTML]{EFEFEF} 
z1              & Null Deref     & Null dereference in z\_nrf\_clock\_control\_lf\_on                                                                                                                                     & \textcolor{blue}{\faUserPlus}                    \\
\rowcolor[HTML]{FFFFFF} 
z2              & OOB Write      & \begin{tabular}[c]{@{}l@{}}The extract\_conversion function in z\_cbvprintf\_impl can cause a potential 1 byte OOB\\ read when the format string ends with a \% character\end{tabular} & \textcolor{blue}{\faUserPlus}                    \\
\rowcolor[HTML]{EFEFEF} 
z2              & OOB Write      & \begin{tabular}[c]{@{}l@{}}If bpe points to a single byte, encode\_uint may cause a 1-byte underflow write by\\ decrementing and dereferencing bp in the loop.\end{tabular}            & \textcolor{blue}{\faUserPlus}                    \\
\rowcolor[HTML]{FFFFFF} 
z2              & OOB Write      & Unchecked length can cause potential overflow                                                                                                                                          & \textcolor{blue}{\faUserPlus}                    \\
\rowcolor[HTML]{EFEFEF} 
z3              & OOB Read       & OOB reads in lv\_txt\_utf8\_next                                                                                                                                                       & \textcolor{green}{\faUserCheck}          \\
\rowcolor[HTML]{FFFFFF} 
z3              & OOB Read       & Reads past the buffer possible in u8\_to\_dec                                                                                                                                          & \textcolor{blue}{\faUserPlus}                    \\
\rowcolor[HTML]{EFEFEF} 
z3              & Div By Zero    & Potential div by zero by passing 0 as data frame size                                                                                                                                  & \textcolor{red}{\faUserTimes}               \\ \midrule
\rowcolor[HTML]{FFFFFF} 
t4              & Div By Zero    & division by zero is possible given RCC-\textgreater{}PLLCFGR is 0                                                                                                                      & \textcolor{green}{\faUserCheck}         \\ \bottomrule
\end{tabular}
\caption{\highlighted{Summary of all reported bugs and their statuses - \mbox{\textcolor{green}{\faUserCheck}} : acknowledged and PR merged, \mbox{\textcolor{blue}{\faUserPlus}} : acknowledged, \mbox{\textcolor{orange}{\faUserClock}} : no response (issue open), \mbox{\textcolor{red}{\faUserTimes}} : not acknowledged as bug and closed.}}
\label{tab:bug_details}
\end{table*}
\begin{table}[]
\small
\centering
\caption{Bugs found by our technique in whole program (C2) and function level (C3) fuzzing modes.}
\vspace{0.25cm}
\begin{tabular}{c|c|c|c}
\toprule
\rowcolor[HTML]{FFFFFF} 
\textbf{App}                                              & \textbf{C2} & \textbf{C3} & \textbf{Unique Bugs} \\ \midrule
\rowcolor[HTML]{EFEFEF} 
f1 & 1 & 1 & \cellcolor[HTML]{FFFFFF}                    \\ \cmidrule{1-3}
f2 & 3 & 1 & \cellcolor[HTML]{FFFFFF}                    \\ \cmidrule{1-3}
\rowcolor[HTML]{FFFFFF} 
\rowcolor[HTML]{EFEFEF}  f3 & 2 & 3 & \cellcolor[HTML]{FFFFFF}                    \\ \cmidrule{1-3}
f4 & 1 & 3 & \cellcolor[HTML]{FFFFFF}                    \\ \cmidrule{1-3}
\rowcolor[HTML]{FFFFFF} 
\rowcolor[HTML]{EFEFEF}  f5 & 2 & 2 & \multirow{-4}{*}{\cellcolor[HTML]{FFFFFF}9} \\ \midrule
n1 & 1 & 1 & \cellcolor[HTML]{FFFFFF}                    \\ \cmidrule{1-3}
n2 & 0 & 1 & \cellcolor[HTML]{FFFFFF}                    \\ \cmidrule{1-3}
\rowcolor[HTML]{FFFFFF} 
\rowcolor[HTML]{EFEFEF}  n3 & 1 & 1 & \cellcolor[HTML]{FFFFFF}                    \\ \cmidrule{1-3}
n4 & 1 & 1 & \cellcolor[HTML]{FFFFFF}                    \\ \cmidrule{1-3}
\rowcolor[HTML]{FFFFFF} 
\rowcolor[HTML]{EFEFEF}  n5 & 1 & 0 & \multirow{-4}{*}{\cellcolor[HTML]{FFFFFF}4} \\ \midrule
z1 & 2 & 4 & \cellcolor[HTML]{FFFFFF}                    \\ \cmidrule{1-3}
\rowcolor[HTML]{FFFFFF} 
\rowcolor[HTML]{EFEFEF}  z2 & 2 & 3 & \cellcolor[HTML]{FFFFFF}                    \\ \cmidrule{1-3}
z3 & 2 & 6 & \multirow{-3}{*}{\cellcolor[HTML]{FFFFFF}7} \\ \midrule
\rowcolor[HTML]{FFFFFF} 
\rowcolor[HTML]{EFEFEF}  t4 & 1 & 1 & \cellcolor[HTML]{FFFFFF}1                                           \\ \midrule
\multicolumn{1}{l|}{\textbf{Total Bugs}} & \textbf{20}        & \textbf{28}             & \textbf{21}          \\

\bottomrule
\end{tabular}
\label{tab:bug_table}
\end{table}
\vspace*{\fill}
\begin{table}[]
\caption{Types of Bugs found by~\systemname{} in both Wp and Fl.}
\vspace{0.25cm}
\centering
\begin{tabular}{ll}
\toprule
\textbf{Bug Type} & \textbf{Count} \\ \midrule
\rowcolor[HTML]{EFEFEF} 
OOB Read          & 2              \\ \midrule
OOB Write         & 7              \\ \midrule
\rowcolor[HTML]{EFEFEF} 
Div By Zero       & 2              \\ \midrule
Null Dereference  & 3              \\ \midrule
\rowcolor[HTML]{EFEFEF} 
DoS               & 2              \\ \midrule
Assert Failure    & 2              \\ \midrule
\rowcolor[HTML]{EFEFEF} 
Build Related     & 3              \\
\bottomrule
\end{tabular}
\label{tab:bug_types}
\end{table}
\vspace*{\fill}
\begin{table*}[]
\centering
\begin{tabular}{cc|c|c}
\toprule
\multicolumn{1}{c|}{\textbf{RTOS}}              & \textbf{CATEGORY}                            & \textbf{Low Fidelity}     & \textbf{High Fidelity}    \\ \toprule
\multicolumn{1}{c|}{}                           & \cellcolor[HTML]{EFEFEF}BOF                  & \cellcolor[HTML]{EFEFEF}3 & \cellcolor[HTML]{EFEFEF}0 \\ \cline{2-4} 
\multicolumn{1}{c|}{}                           & OOB                                          & 5                         & 0                         \\ \cline{2-4} 
\multicolumn{1}{c|}{}                           & \cellcolor[HTML]{EFEFEF}UAF                  & \cellcolor[HTML]{EFEFEF}4 & \cellcolor[HTML]{EFEFEF}0 \\ \cline{2-4} 
\multicolumn{1}{c|}{}                           & Int Overflow                                 & 8                         & 1                         \\ \cline{2-4} 
\multicolumn{1}{c|}{\multirow{-5}{*}{FreeRTOS}} & \cellcolor[HTML]{EFEFEF}Privelege Escalation & \cellcolor[HTML]{EFEFEF}0 & \cellcolor[HTML]{EFEFEF}1 \\ \midrule
\multicolumn{1}{c|}{}                           & BOF                                          & 10                        & 2                         \\ \cline{2-4} 
\multicolumn{1}{c|}{}                           & OOB                                          & 5                         & 1                         \\ \cline{2-4} 
\multicolumn{1}{c|}{}                           & \cellcolor[HTML]{EFEFEF}Int Overflow         & \cellcolor[HTML]{EFEFEF}3 & \cellcolor[HTML]{EFEFEF}1 \\ \cline{2-4} 
\multicolumn{1}{c|}{}                           & DoS                                          & 4                         & 1                         \\ \cline{2-4} 
\multicolumn{1}{c|}{}                           & \cellcolor[HTML]{EFEFEF}Privelege Escalation & \cellcolor[HTML]{EFEFEF}0 & \cellcolor[HTML]{EFEFEF}2 \\ \cline{2-4} 
\multicolumn{1}{c|}{\multirow{-6}{*}{Zephyr}}   & NULL Deref                                   & 4                         & 0                         \\ \midrule
\multicolumn{1}{c|}{}                           & \cellcolor[HTML]{EFEFEF}BOF                  & \cellcolor[HTML]{EFEFEF}2 & \cellcolor[HTML]{EFEFEF}1 \\ \cline{2-4} 
\multicolumn{1}{c|}{}                           & OOB                                          & 3                         & 0                         \\ \cline{2-4} 
\multicolumn{1}{c|}{}                           & \cellcolor[HTML]{EFEFEF}NULL Deref           & \cellcolor[HTML]{EFEFEF}5 & \cellcolor[HTML]{EFEFEF}0 \\ \cline{2-4} 
\multicolumn{1}{c|}{}                           & DOS                                          & 2                         & 1                         \\ \cline{2-4} 
\multicolumn{1}{c|}{\multirow{-5}{*}{RIOT}}     & \cellcolor[HTML]{EFEFEF}logic                & \cellcolor[HTML]{EFEFEF}2 & \cellcolor[HTML]{EFEFEF}0 \\ \midrule
\multicolumn{2}{c|}{\textbf{Total}}                                                            & \textbf{60}               & \textbf{11}               \\ \bottomrule
\end{tabular}
\caption{\highlighted{Our analysis of the CVEs from the survey\mbox{~\cite{Rust4EmbeddedSurvey}} conducted by Rust4Embedded (Extended Report)\mbox{~\cite{sharma2023rust}} indicates that only 11 out of 71 (15\%) require high-fidelity execution (i.e precise hardware modeling).}}
\label{tab:embedded_bug_survey}
\end{table*}

\subsection{Implementation Details}

\begin{listing}
    \begin{minted}[xleftmargin=0.25cm, numbersep=1pt, escapeinside=||, fontsize=\scriptsize, breaklines, linenos]{cpp}

    |\faFileCode| Effected Source file
    struct foo {
           int x;
          int y[];
    };
    struct foo bar = {1, {2, 3, 4}};

    |\textcolor{red}{\faCalendarTimes}| Compilation Fails
    error: initialization of flexible array member is not allowed
    struct foo bar = {1, {2, 3, 4}};

    |\faChalkboardTeacher| Lemix Provides the Following Guidance Instruction
    
    1. Search for the definition of the structure which has the
        flexible member.
    2. The member declaration will have a [], add a constant
        value ex. [100] to the member.
    3. Re-run the framework.
  
\end{minted}
\caption{\faFileCode\ shows a sample code which causes compilation error indicated by \textcolor{red}{\faCalendarTimes}.~\systemname{} provides instructions to the developer to assist in fixing errors shown by \faChalkboardTeacher.}
\label{lst:lemix_guidance}
\end{listing}

We use a combination of Python scripts and \clang{}/\llvm{} 10 toolchain passes to implement our framework.
\subsection{Phase 1: \elapp{} generation and Instrumentation}
\label{apdx:subsecimpl}
We implement our \elapp{} generation as a Python tool.
Given an embedded application, we first build for one of the supported targets and capture the compilation and linking steps using the Build EAR (Bear) tool \cite{bear} in a JSON file, \ie{} \inlinecode{compile_commands.json}.
We use \wllvm{} tool \cite{wllvmtool} to generate bitcode files for each of the source files.
As mentioned in \sect{subsubsec:interactivecompilererrorresolution} and \tbl{tab:gcc_clang_diffs}, our tool uses an interactive technique to resolve compiler incompatibilities.
Next, given the target \ac{RTOS} for the embedded application, we identify the corresponding source files and replace them with the bitcode files of \ac{LPL} layer to produce the final \elapp{}
We will tackle the linker issues (as mentioned in \sect{subsubsec:oppsymbolresolution}), through a set of binary analysis scripts and \llvm{} passes.

We also use \llvm{} transformation passes to implement our peripheral modeling and dynamic analysis assistance.
We created a runtime library (one-time effort) implementing our hooks (\eg{} \inlinecode{get_input_from_stdin()} in \lst{lst:mmio_instrumentation}), which will be linked to produce the final fuzz-ready \elapp{}.

\subsection{Phase 2: Testing}
As mentioned in \sect{subsec:types_emboss}, \elapp{} runs indefinitely in an event-driven mode.
For whole program fuzzing, we use the persistent mode of \afl{} to provide inputs in a continuous manner and record coverage at specific intervals.
We have also created auxiliary scripts to assist in crash reproducibility.

For function level fuzzing, we use the recent {\sc 3c} tool \cite{machiry2022c} for our co-relation analysis and implement our generator creation and \elapp{} modification as \llvm{} passes.

\section{Appendix}
\label{apdx:appendix_c}

\subsection{Dataset Details}
\tbl{tab:full_dataset_table} provides details of all applications used in our evaluation.

\subsection{Bugs Found By~\systemname{}}

\tbl{tab:bug_table} shows bugs found by \systemname{} across all the RTOSes and \tbl{tab:bug_types} shows the types of bugs found. \tbl{tab:bug_details} gives a detailed description and developer responses for each bug found.

\section{Appendix}
\label{apdx:appendix_d}

\begin{listing}[tb]
    \begin{minted}[xleftmargin=0.25cm, numbersep=1pt, escapeinside=||, fontsize=\scriptsize, breaklines, linenos]{C}
    
    __STATIC_FORCEINLINE uint32_t __get_IPSR(void) 
    {
        uint32_t result;
        |\faCogs|__ASM volatile ("MRS %0, ipsr" : "=r" (result));
        |\textcolor{blue}{\faPlusSquare}| result = random() % 2; |\textcolor{blue}{\faPlusSquare}|
        return result;
    }
    #define FURI_IS_IRQ_MODE() ({__get_IPSR() != 0}) |\textcolor{green}{\faCheckSquare}|
    
    bool furi_kernel_is_irq_or_masked(void) {
        return {FURI_IS_IRQ_MODE()};}
    int main(void) {
        // furi_check handles assertions
        furi_check(
           {!furi_kernel_is_irq_or_masked()} |\textcolor{green}{\faCheckSquare}|
        );
        return 0;
    }
    \end{minted}
    \caption{An example of inline assembly effecting initialization code in Flipperzero. The \faCogs\ indicates inline assembly code removed by~\systemname{} and \textcolor{blue}{\faPlusSquare} indicates injected code to re-initialized with a random value. The highlighted checks (\textcolor{green}{\faCheckSquare}) show how the IPSR value is verified in the initialization process to ensure we're not in IRQ}
    \label{lst:inlineASM}
\end{listing}
\begin{listing}[tb]
    \begin{minted}[xleftmargin=0.25cm, numbersep=1pt, fontsize=\scriptsize, breaklines, linenos]{C}
    __ALIGN(16)
    static const uint16_t delay_machine_code[] = {
        // SUBS r0, #loop_cycles
        0x3800 + NRFX_COREDEP_DELAY_US_LOOP_CYCLES, 
        0xd8fd, // BHI .-2
        0x4770  // BX LR
    };

    typedef void (* delay_func_t)(uint32_t);
    const delay_func_t delay_cycles =
        // Set LSB to 1 to execute the code in the 
        // Thumb mode.
        (delay_func_t)(delay_machine_code) | 1));
    uint32_t cycles = 
        time_us * NRFX_DELAY_CPU_FREQ_MHZ;
    delay_cycles(cycles);
    \end{minted}
    \caption{An example of inline assembly masked inside hexadecimal machine code in Infinitime, one of our FreeRTOS applications.}
    \label{lst:hexASM}
\end{listing}
\begin{listing}
    \begin{minted}[xleftmargin=0.25cm, numbersep=1pt, escapeinside=||, fontsize=\scriptsize, breaklines, linenos]{cpp}
    int main() {
        /* RTOS specific task creation */
        xTaskCreate(cdc_task, "cdc", CDC_STACK_SIZE, NULL,
            configMAX_PRIORITIES - 2, NULL);
        ...
        /* Injecting isr_trigger function as a task
        along with other tasks */
        xTaskCreate(isr_trigger, "dispatcher_task", 1000, NULL,
            configMAX_PRIORITIES - 1, NULL);
    }
    
    void dispatcher_task(void *parm) {
       (void)parm;
       while(1) {
            int size = sizeof(ivt) / sizeof(ivt[0]);
            int random_isr = rand() % size;
            switch(random_isr) {
                case 0:
                    USBD_IRQHandler();
                    break;
                ...
                case 4:
                    SPIM1_TWI1_IRQHandler();
                    break;
                default:
                    break;
            } } }

\end{minted}
\caption{Dispatcher Task instrumentation to handle interrupts.}
\label{lst:isr_template}
\end{listing}
\begin{listing}
    \begin{minted}[xleftmargin=0.25cm, numbersep=1pt, escapeinside=||, fontsize=\scriptsize, breaklines, linenos]{cpp}

    static void z_sys_init_run_level(
        enum init_level level
    )
    {
       static const struct init_entry *levels[] = {
           __init_EARLY_start,
           __init_PRE_KERNEL_1_start,
           __init_PRE_KERNEL_2_start,
           __init_POST_KERNEL_start,
           __init_APPLICATION_start,
       };
       const struct init_entry *entry;
       // The entries are function pointers that
       // are expected to be placed in memory in
       // the correct order. This ensures that 
       // the comparison between the current entry
       // and the next one is valid.
       for (entry = levels[level]; entry < 
            levels[level+1]; entry++) {
               dev->state->initialized = true;
                (void)entry->init_fn.sys();
           }
       }
    }   
\end{minted}
\caption{Listing shows Layout Specific Execution found in one of Zephyr Kernel's initialization routines. The kernel expects function pointers to be present in adjacent memory locations as directed by Board Specific Linker Scripts.}
\label{lst:board_layout}
\end{listing}
\begin{listing}
    \begin{minted}[xleftmargin=0.25cm, numbersep=1pt, escapeinside=||, fontsize=\scriptsize, breaklines, linenos]{cpp}
    
    void isr_trigger(void *parm) {
        void isr_trigger(void *parm) {
       (void)parm;
       while(1) {
            int size = sizeof(ivt) / sizeof(ivt[0]);
            int random_isr = rand() % size;
            switch(random_isr) {
                case ...
                case 4:
                    /* ISR expecting a global state to be 
                    initialized */
                    SPIM1_TWI1_IRQHandler(); |\textcolor{red}{\faExclamationCircle}|
                    break;
                default:
                    break;
            }
        }
    }

    |\textcolor{red}{\faExclamationTriangle}|
    void SPIM1_TWI1_IRQHandler(void)
    {
        /* ISR expecting a global state m_cb
         *  to be initialized. 
         */
        irq_handler(((NRF_SPIM_Type*) 0x40004000UL),
            &m_cb[NRFX_SPIM1_INST_IDX]);
    }
    |\textcolor{red}{\faExclamationTriangle}|

    |\faCogs|
    void board_init(void) {
    /* Everything necessary gets initialized here */

    #if CFG_TUH_ENABLED && defined(CFG_TUH_MAX3421) \
        && CFG_TUH_MAX3421
        /* SPIM1_TWI1_IRQHandler initialized only when 
         * any of these configs are selected.
         */
        
        max3421_init(); |\textcolor{black}{\faFrownOpen}|
    #endif
    }
    |\faCogs|

\end{minted}
\caption{\highlighted{An example of a false positive interrupt misfiring in TinyUSB Application.}}
\label{lst:isr_fp}
\end{listing}

\lst{lst:inlineASM} shows an example of inline ASM affecting program initialization where as \lst{lst:hexASM} shows an example of machine code embedded into the application as hex codes which is challenging to analyze on source level.\lst{lst:isr_template} demonstrates how ISRs are modelled by~\systemname{} to run alongside other application tasks and \lst{lst:board_layout} gives an example of kernel initialization code from Zephyr RTOS application exhibiting memory layout dependent code which is currently not supported by~\systemname{}.

\section{Appendix}
\label{apdx:appendix_e}

\begin{table*}[]
\caption{
Table depicting common GCC vs Clang incompatibilities found while porting applications across various RTOSes.
}
\vspace{0.25cm}
\scriptsize
\centering
\renewcommand{\arraystretch}{1}

\newcolumntype{C}[1]{>{\centering\arraybackslash}m{#1}}
\begin{tabularx}{\textwidth}{C{3cm}|C{2cm}|C{1.75cm}|C{2cm}|C{2cm}|C{3cm}|C{1cm}}
\toprule
\textbf{Error Type} & \multicolumn{4}{c|}{\textbf{Compiler Support}} & \textbf{Error Fix} & \textbf{Automated} \\ \cmidrule{2-5}
 & \textbf{GCC Specific} & \textbf{Arm-GCC Specific} & \textbf{Architecture Specific} & \textbf{Clang Support} & \\ 
 \midrule
Initialization of flexible array member is not allowed. & Yes (Example~\ref{lst:flex_init}) & No & No & This initialization is not allowed in clang (<15). & Make the array member of a fixed size. & No \\ \midrule
Variable-sized object may not be initialized. & Yes (Example~\ref{lst:var_size_obj}) & No & No & No & Declare the local variable with a specified length before initializing it. & Yes \\ \midrule
Member initializer 'X' does not name a non-static data member or base class. & Yes (Example~\ref{lst:class_glob_space}) & No & No & No & Make the class inherit from global namespace. & No \\ \midrule
Static Assertion Failed. & No & No & This is due to architecture specific size of types. Assert failure is due to compiling 32 bit compatible types on a 64 bit machine. & NA & Compile target on 32 bit equivalent architecture. & Yes \\ \midrule
Undeclared identifier example, \textit{\_\_assert\_func} & No & Yes, The symbol is specific to ARM's internal headers. & No & No & Get the symbol definition from the original source's preprocessor output. & Semi-Automated \\ 
\bottomrule
\end{tabularx}
\vspace{0.25cm}
\label{tab:gcc_clang_diffs}
\end{table*}
\begin{listing}
    \begin{minted}[xleftmargin=0.25cm, numbersep=1pt, escapeinside=||, fontsize=\scriptsize, breaklines, linenos]{cpp}
    int
    doread(uint8_t addr, uint8_t *buf, uint8_t len)
    {
        uint8_t tx_buf[len + 1] = addr;
        return  -1;
    }

\end{minted}
\caption{Listing shows variable sized object initialization which is not allowed even in the latest version of clang.}
\label{lst:var_size_obj}
\end{listing}
\begin{listing}
    \begin{minted}[xleftmargin=0.25cm, numbersep=1pt, escapeinside=||, fontsize=\scriptsize, breaklines, linenos]{cpp}
    // Base class representing a generic object
    class b {
    public:
        // Constructor accepting a config
        b(const int config);
        // Virtual destructor
        virtual ~b() = default;  
    };
    
    namespace a::b::c {
    
    class b_c : public b {
    public:
        // Constructor for b_c, initializes base class b
        b_c(const int config);  
    };
    
    // Constructor implementation for b_c
    b_c::b_c(const int config) : b(config) {}
    
    }

\end{minted}
\caption{Listing shows a sample code which errors because Clang expects the class b to inherit from the global namespace, but it cannot find it within the current scope.}
\label{lst:class_glob_space}
\end{listing}
\begin{listing}
    \begin{minted}[xleftmargin=0.25cm, numbersep=1pt, escapeinside=||, fontsize=\scriptsize, breaklines, linenos]{cpp}
     struct foo {
           int x;
          int y[];
  };

  struct foo bar = {1, {2, 3, 4}};

\end{minted}
\caption{Listing shows flexible array member initialization which is not allowed in clang versions < 15.}
\label{lst:flex_init}
\end{listing}

\subsection{GCC vs Clang Differences}

\tbl{tab:gcc_clang_diffs} provides details on the compiler incompatibilities. Listings~\ref{lst:flex_init},~\ref{lst:var_size_obj} and~\ref{lst:class_glob_space} show examples of code that fails to compile with clang but is compatible with GCC.

\section{Appendix}
\label{apdx:appendix_f}
\begin{listing}[tb]
    \begin{minted}[xleftmargin=0.25cm, numbersep=1pt, escapeinside=||, fontsize=\scriptsize, linenos]{c}
    uint8_t ull_scan_rsp_set(struct ll_adv_set *adv,
    uint8_t len, void *data)
    {
        struct pdu_adv *pdu;
        pdu = lll_adv_scan_rsp_alloc(&adv->lll, &idx);
        /* update scan pdu fields. */
        ...
        /* len is attacker controlled */
        pdu->len = BDADDR_SIZE + len; |\textcolor{orange}{\faUserSecret}|
        /* OOB write at scan_rsp.data[0] */
        memcpy(&pdu->scan_rsp.data[0], data, len); |\textcolor{red}{\faBug}|
        ...
        return 0;
    }   
    \end{minted}
    \caption{\highlighted{CVE-2021-3581: A low-fidelity vulnerability where unchecked length can cause OOB write if data and len are attacker controlled.}}
    \label{lst:low_fidelity_cve}
\end{listing}
\begin{listing}[tb]
\begin{minted}[xleftmargin=0.25cm, numbersep=1pt, escapeinside=||, fontsize=\scriptsize, breaklines, linenos]{C}
    
    |\textcolor{blue}{\faCheckCircle}| Identified Address ranges that are also in SVD:
    
     (0x40000000, 0x40001000)
    |\faMicrochip| (Peripheral: TIM2, Base: 0x40000000, End: 0x40000400)
    
    (0x40008000, 0x40009000)
    |\faMicrochip| (Peripheral: LPTIM1, Base: 0x40007c00, End: 0x40008000)
    
    (0x40010000, 0x40011000)
    |\faMicrochip| (Peripheral: SYSCFG_VREFBUF, Base: 0x40010000, End: 0x40010200)
    
    (0x40020000, 0x40021000)
    |\faMicrochip| (Peripheral: DMA1, Base: 0x40020000, End: 0x40020400)
    
    |\textcolor{red}{\faExclamationTriangle}| Identified Address ranges (and corresponding source lines) that are not in SVD files:

    #define PERIPH_BASE           (0x40000000UL)
    #define APB1PERIPH_BASE       PERIPH_BASE
    #define TIM16_BASE            (APB2PERIPH_BASE + 0x00004400UL)
    (0x40020000, 0x40021000)
    
    #define SAI1_BASE             (APB2PERIPH_BASE + 0x00005400UL)
    (0x40005000, 0x40006000)
    
    #define LPTIM2_BASE           (APB1PERIPH_BASE + 0x00009400UL)
    (0x40009000, 0x4000a000)
\end{minted}
\caption{An example illustrating the cases where the detected MMIO address ranges are in SVD files (\faMicrochip) and those that are present in source files but missing (\textcolor{red}{\faExclamationTriangle}) in SVD files.}
\label{lst:mmio_extra}
\end{listing}
\begin{listing}
    \begin{minted}[xleftmargin=0.25cm, numbersep=1pt, escapeinside=||, fontsize=\scriptsize, breaklines, linenos]{cpp}

    bool latch_pending_read_and_check(latch, i) {
        latch[i] = GPIOx->LATCH; |\textcolor{orange}{\faUserSecret}|
        if (latch[i])
        {
            /* If any latch bit is set, another edge was
            captured — repeat event processing. */
            return true;
        }
    }
    void port_event_handle(...) 
    {
        uint32_t latch[GPIO_COUNT] = {0};
        ...
        do {
            /* Attacker controlled gpio ports */
            latch_pending_read_and_check(latch,
                            p_cb->available_gpio_ports);
            /* The latch sent for processing */
            nrfy_gpiote_events_process(p_gpiote, ...);

        /* The latch_pending_read_and_check can always
            return true leading to infinite loop */
        } while (latch_pending_read_and_check(latch,  |\textcolor{red}{\faBug}|
            p_cb->available_gpio_ports));
    }
    
\end{minted}
\caption{\highlighted{If \mbox{\inlinecode{latch_pending_read_and_check()}} keeps returning true due to malicious GPIO peripheral, the do-while loop will never terminate, causing an infinite loop leading to DoS found in nrfx HAL library.}}
\label{lst:mmio_infinite_loop}
\end{listing}
\setlength{\tabcolsep}{4pt}
\begin{table*}
 \caption{Approximate Number of unique basic blocks discovered by various configurations of~\systemname{} in comparison to State Of The Art Tools (Discussed in \sect{subsec:testingelapp} and \sect{subsec:comparativeeval}).
 M1- M3 represents different configuration modes for~\systemname{}.}
\vspace{0.25cm}
\centering
\rowcolors{5}{gray!15}{white}
\begin{tabular}{c | c c | c c | c c | c c c | c c c}
\toprule
& \multicolumn{6}{c|}{\textbf{Lx}} & \multicolumn{3}{c|}{\multirow{2}{*}{\textbf{Fw}}} & \multicolumn{3}{c}{\multirow{2}{*}{\textbf{Mf}}} \\
\cmidrule{2-7}
\multirow{-3}{*}{\textbf{AppID}} & 
\multicolumn{2}{c|}{\textbf{M1}} & \multicolumn{2}{c|}{\textbf{M2}} & \multicolumn{2}{c|}{\textbf{M3}} & 
\multicolumn{3}{c|}{\textbf{}} & 
\multicolumn{3}{c}{\textbf{}} \\ 
\cmidrule{2-7} \cmidrule{8-13}
& Cov & Bug & Cov & Bug & Cov & Bug & Cov & Crash & Bug & Cov & Crash & Bug \\
\midrule
f1 & 731 & 0 & 2.9k & 1 & \textcolor{red}{\faCode} & 1 & 500 & 1 & 0 & 1k & 0 & 0 \\
\hline
f2 & 456 & 1 & 2.8k & 3 & \textcolor{red}{\faCode} & 1 & \textcolor{orange}{\faExclamationCircle} & N/A & N/A & 1.5k & 1 & 1 \\ 
\hline
f3 & 560 & 1 & 668 & 2 & 1.5k & 3 & \textcolor{orange}{\faExclamationCircle} & N/A & N/A & \textcolor{orange}{\faExclamationCircle} & N/A & N/A \\
\hline
f4 & 563 & 0 & 1k & 1 & 6k & 3 & 500 & 0 & 0 & 2k & 41 & 0 \\
\hline
f5 & 442 & 0 & 728 & 2 & 1.8k & 2 & 700 & 0 & 0 & 1.8k & 93 & 0 \\
\midrule
n1 & 105 & 0 & 301 & 1 & 13.5k & 1 & 356 & 0 & 0 & 25.2k & 148 & 0 \\
\hline
n2 & 143 & 0 & 338 & 0 & 16.8k & 1 & 405 & 0 & 0 & 300 & 0 & 0 \\
\hline
n3 & 157 & 1 & 357 & 1 & 19.9k & 1 & 60 & 1 & 1 & 100 & 1 & 0 \\
\hline
n4 & 135 & 0 & 235 & 1 & 19.5k & 1 & 388 & 3 & 0 & 2.4k & 0 & 0 \\
\hline
n5 & 823  & 0 & 1.5k & 1 & \textcolor{red}{\faCode} & 0 & 134 & 0 & 0 & 226 & 0 & 0 \\
\midrule
z1 & 76 & 0 & 86 & 2 & 3k & 4 & 44 & 2 & 0 & 213 & 0 & 0 \\
\hline
z2 & \textcolor{blue}{\faMicrochip} & 0 & \textcolor{blue}{\faMicrochip} & 2 & 12.3k & 3 & \textcolor{orange}{\faExclamationCircle} & N/A & N/A & 10.8k &  483 & 1 \\
\hline
z3 & \textcolor{blue}{\faMicrochip} & 0 & \textcolor{blue}{\faMicrochip} & 2 & 4.6k & 6 & \textcolor{orange}{\faExclamationCircle} & N/A & N/A & \textcolor{orange}{\faExclamationCircle} & N/A & N/A \\
\midrule
t1 & 210 & 0 & 553 & 0 & 6.3k & 0 & 670 & 0 & 0 & 750 & 0 & 0 \\
\hline
t2 & 214 & 0 & 553 & 0 & 240 & 0 & 791 & 0 & 0 & 694 & 0 & 0 \\
\hline
t3 & \textcolor{blue}{\faMicrochip} & 0 & \textcolor{blue}{\faMicrochip} & 0 & 3.1k & 0 & 900 & 0 & 0 & 700 & 0 & 0 \\
\hline
t4 & 310 & 0 & 455 & 1 & 188 & 1 & 749 & 0 & 0 & 659 & 10 & 1 \\
\hline
t5 & 254 & 0 & 436 & 0 & 5.2k & 0 & 748 & 0 & 0 & 658 & 0 & 0 \\ \midrule
Avg/Tot & 345 & 3 & 860 & 20 & 7.5k & 28 & 496  & 7 & 1 & 3.1k & 777 & 3 \\ 
\midrule
Unique Bugs &  & 3 &  & 10 &  & 11 &  &  & 1 &  &  & 3 \\ 

\bottomrule
\end{tabular}
\label{tab:tool_eval_large}
\end{table*}
\definecolor{bblue}{RGB}{31, 119, 180}   
\definecolor{rred}{RGB}{214, 39, 40}    
\definecolor{ggreen}{RGB}{44, 160, 44}  
\definecolor{ppurple}{RGB}{148, 103, 189} 
\definecolor{oorange}{RGB}{255, 127, 14} 
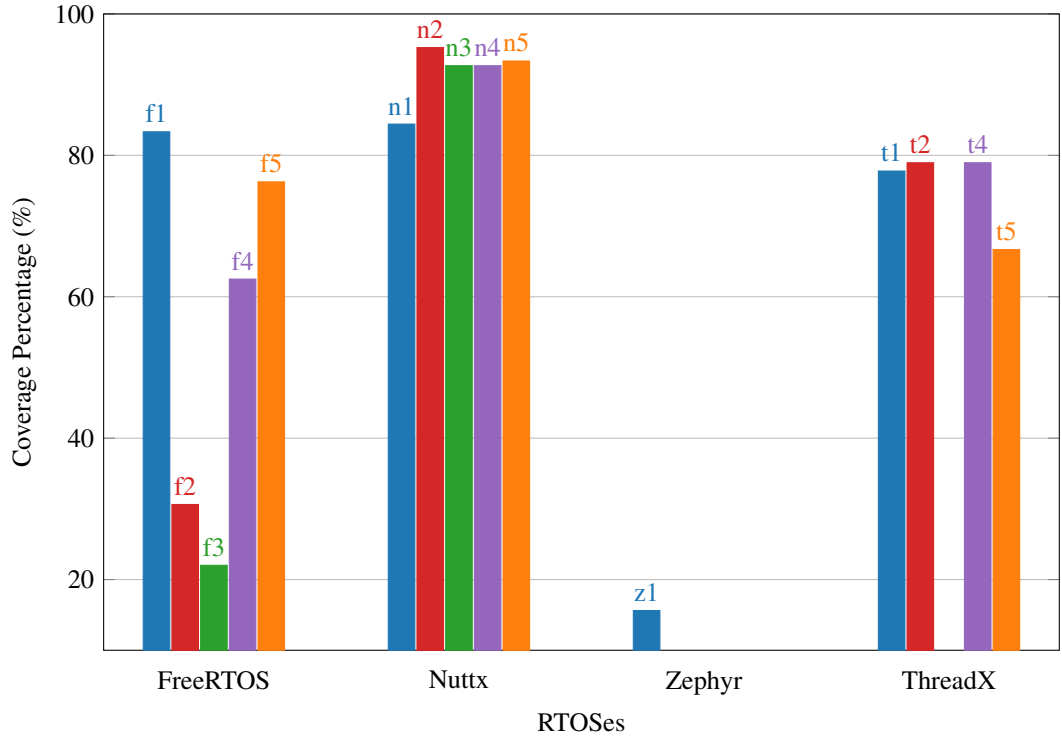
\begin{figure*}[h!]
    \centering
    \begin{tikzpicture}
        \begin{axis}[
            width  = 0.8*\textwidth,
            height = 10cm,
            major x tick style = transparent,
            ybar=2*\pgflinewidth,
            bar width=10pt,
            ymajorgrids = true,
            ylabel = {Coverage Percentage (\%)},
            xlabel = {RTOSes},
            symbolic x coords={FreeRTOS, Nuttx, Zephyr, ThreadX},
            xtick = data,
            scaled y ticks = false,
            enlarge x limits=0.15,
            ymin=10,
            ymax=100,
        ]
        \addplot[style={bblue,fill=bblue,mark=none},
            nodes near coords={
                \ifnum\coordindex=0 f1
                \else\ifnum\coordindex=1 n1
                \else\ifnum\coordindex=2 z1
                \else t1\fi\fi\fi
            },
            nodes near coords style={text=bblue, anchor=south}]
            coordinates {(FreeRTOS, 83.33) (Nuttx, 84.42) (Zephyr, 15.62) (ThreadX, 77.78)};
        
        \addplot[style={rred,fill=rred,mark=none},
            nodes near coords={
                \ifnum\coordindex=0 f2
                \else\ifnum\coordindex=1 n2
                \else t2\fi\fi
            },
            nodes near coords style={text=rred, anchor=south}]
            coordinates {(FreeRTOS, 30.62) (Nuttx, 95.24) (ThreadX, 78.95)};
        
        \addplot[style={ggreen,fill=ggreen,mark=none},
            nodes near coords={
                \ifnum\coordindex=0 f3
                \else n3\fi
            },
            nodes near coords style={text=ggreen, anchor=south}]
            coordinates {(FreeRTOS, 22.02) (Nuttx, 92.68)};
        
        \addplot[style={ppurple,fill=ppurple,mark=none},
            nodes near coords={
                \ifnum\coordindex=0 f4
                \else\ifnum\coordindex=1 n4
                \else t4\fi\fi
            },
            nodes near coords style={text=ppurple, anchor=south}]
            coordinates {(FreeRTOS, 62.50) (Nuttx, 92.68) (ThreadX, 78.95)};
        
        \addplot[style={oorange,fill=oorange,mark=none},
            nodes near coords={
                \ifnum\coordindex=0 f5
                \else\ifnum\coordindex=1 n5
                \else t5\fi\fi
            },
            nodes near coords style={text=oorange, anchor=south}]
            coordinates {(FreeRTOS, 76.25) (Nuttx, 93.34) (ThreadX, 66.67)};
            
        \end{axis}
    \end{tikzpicture}
    \caption{Each bar represents the percentage of functions hit amongst the total reachable functions for an application on a specific RTOS. Application IDs represent respective application from Table~\ref{tab:full_dataset_table}.}
    \label{fig:function_coverage}
\end{figure*}
\begin{listing}[tb]
    \begin{minted}[xleftmargin=0.25cm, numbersep=1pt, escapeinside=||, fontsize=\scriptsize, breaklines, linenos]{gas}

     /* validate syscall limit */
    ldr ip, =K_SYSCALL_LIMIT
    cmp r6, ip
    /* The supplied syscall_id must be lower than the 
     *  limit (Requires unsigned integer comparison)
     */
    blt valid_syscall_id |\textcolor{red}{\faBug}|

    /* bad syscall id.  Set arg1 to bad id and set
        call_id to SYSCALL_BAD */
    str r6, [r0]
    ldr r6, =K_SYSCALL_BAD
    \end{minted}
    \caption{\highlighted{CVE-2020-10027: A high-fidelity vulnerability where a signed comparison in ARM syscall validation (blt vs. blo) allows privilege escalation from user thread to kernel.}}
    \label{lst:high_fidelity_cve}
\end{listing}

\lst{lst:mmio_extra} gives an example of a valid MMIO address range detected by~\systemname{}'s constant address analysis which was not found in the corresponding SVD files. \tbl{tab:tool_eval_large} provides a more readable version of evaluation of~\systemname{} against existing state of the art tools.\fig{fig:function_coverage} gives an idea of the coverage of total reachable functions in whole-program fuzzing.

\subsection{Bug Detected in Zephr RTOS}
\label{apdx:bug_detection}
\begin{listing}
\begin{minted}[xleftmargin=0.25cm, numbersep=1pt, escapeinside=||, fontsize=\scriptsize, breaklines, linenos]{cpp}
    #ifndef CONFIG_PRINTK_BUFFER_SIZE
    #define CONFIG_PRINTK_BUFFER_SIZE 0 |\textcolor{yellow}{\faExclamationTriangle}|
    struct buf_out_context {
        char buf[CONFIG_PRINTK_BUFFER_SIZE];
        unsigned int buf_count;
    };
    
    static int buf_char_out(int c, void *ctx_p) {
        struct buf_out_context *ctx = ctx_p;
        ctx->buf[ctx->buf_count] = c; |\textcolor{red}{\faBug}|
        // buf_count incremented before the check
        ++ctx->buf_count; |\textcolor{blue}{\faSkull}|
        if (ctx->buf_count == CONFIG_PRINTK_BUFFER_SIZE) {
            buf_flush(ctx);
        }
        return c;
    }
\end{minted}
\caption{OOB write in Zephyr function \textit{buf\_char\_out}.}
\label{lst:identified_bugs_1}
\end{listing}
\lst{lst:identified_bugs_1} shows a buffer out-of-bounds write bug in Zephyr \ac{RTOS}, where if \textit{CONFIG\_PRINTK\_BUFFER\_SIZE} is not set by the application, it defaults to 0.
At the \textit{cbvprintf} call site, the \textit{buf\_count} member of \textit{ctx} is initialized to 0, leading to out-of-bounds access in the \textit{buf\_char\_out} function.
This bug occurs in all applications of Zephr.



\end{document}